\documentclass[twocolumn,showkeys,preprintnumbers,amsmath,amssymb,prb,floatfix,superscriptaddress,longbibliography]{revtex4-1}
\usepackage{graphicx,tabularx}
\usepackage[version=3]{mhchem} % Formula subscripts using \ce{}
\usepackage{amssymb}
\usepackage{dcolumn}
\usepackage{color}% to create grayscales for the ball and stick images
\usepackage[mathcal]{euscript}
\usepackage{color}
\usepackage{siunitx} %\shannon{I added siunitx}
\definecolor{gray0}{gray}{0.0}%black
\definecolor{gray64}{gray}{0.25}
\definecolor{gray128}{gray}{0.5}
\definecolor{gray192}{gray}{0.75}
\definecolor{gray255}{gray}{1.0}%white

\begin{document}
This work was published in\\
Diamond and Related Materials (2019).\\
Doi: 10.1016/j.diamond.2019.02.024

%\title{Eu-doping of Diamond: The Quest for Luminescent Centres Through a Combined Theoretical--Experimental Study}
\title{Can Europium Atoms form Luminescent Centres in Diamond: A combined Theoretical--Experimental Study}
\author{Danny E. P. Vanpoucke}
\email{Danny.Vanpoucke@UHasselt.be}
\affiliation{Institute for Materials Research (IMO), Hasselt University, 3590 Diepenbeek, Belgium}
\affiliation{IMOMEC, IMEC vzw, 3590 Diepenbeek, Belgium}
\author{Shannon S. Nicley}
\affiliation{Institute for Materials Research (IMO), Hasselt University, 3590 Diepenbeek, Belgium}
\affiliation{IMOMEC, IMEC vzw, 3590 Diepenbeek, Belgium}
\affiliation{Department of Materials, University of Oxford, Oxford, UK}
\author{Jorne Raymakers}
\affiliation{Institute for Materials Research (IMO), Hasselt University, 3590 Diepenbeek, Belgium}
\affiliation{IMOMEC, IMEC vzw, 3590 Diepenbeek, Belgium}
\author{Wouter Maes}
\affiliation{Institute for Materials Research (IMO), Hasselt University, 3590 Diepenbeek, Belgium}
\affiliation{IMOMEC, IMEC vzw, 3590 Diepenbeek, Belgium}
\author{Ken Haenen}
\email{Ken.Haenen@UHasselt.be}
\affiliation{Institute for Materials Research (IMO), Hasselt University, 3590 Diepenbeek, Belgium}
\affiliation{IMOMEC, IMEC vzw, 3590 Diepenbeek, Belgium}

\date{\today}
\begin{abstract}
The incorporation of Eu into the diamond lattice is investigated in a combined theoretical--experimental study. The large size of the Eu ion induces a strain on the host lattice, which is minimal for the Eu-vacancy complex. The oxidation state of Eu is calculated to be $3+$ for all defect models considered. In contrast, the total charge of the defect-complexes is shown to be negative: $-1.5$ to $-2.3$ electron. Hybrid-functional electronic-band-structures show the luminescence of the Eu defect to be strongly dependent on the local defect geometry. The 4-coordinated Eu substitutional dopant is the most promising candidate to present the typical Eu$^{3+}$ luminescence, while the 6-coordinated Eu-vacancy complex is expected not to present any luminescent behaviour. Preliminary experimental results on the treatment of diamond films with Eu-containing precursor indicate the possible incorporation of Eu into diamond films treated by drop-casting. Changes in the PL spectrum, with the main luminescent peak shifting from approximately $614$ nm to $611$ nm after the growth plasma exposure, and the appearance of a shoulder peak at $625$ nm indicate the potential incorporation. Drop-casting treatment with an electronegative polymer material was shown not to be necessary to observe the Eu signature following the plasma exposure, and increased the background luminescence.
\end{abstract}
% need to include ge-review paper on NW's

\keywords{Diamond, Eu, doping, defects, DFT, Density Functional Theory, electronic structure, phonons, photoluminescence}
%\pacs{  }
%----------------------------------------------------------------
\maketitle
%------------------------------------------------------------------------------------------
%------------------------------Introduction------------------------------------------------
%------------------------------------------------------------------------------------------
\section{Introduction}
\indent Many of the current day quantum technology applications make use of diamond, more specifically diamond with nitrogen-vacancy (NV) defects.\cite{ChildressH:MRSBull2013,BougasL:Micromachines2018} This optically active defect makes it possible for diamond to be used for high resolution magnetometery, quantum entanglement experiments, such as testing of the Bell inequality, and much more.\cite{AharonovichI:AdvOptMater2014, PfaffW:Science2014, HensenB:Nature2015, CasolaF:NatRevMater2018, ThieringG:PhysRevB2018, UdvarhelyiP:PhysRevB2018} Although it is possible to produce NV centres in diamond in a controlled way, the efficiency of the NV centre itself is limited by its strong electron-phonon coupling.\cite{AharonovichI:AdvOptMater2014,PlakhotnikT:PhysRevB2015} However, because diamond has a very large band gap this provides opportunities for the use of alternate optically active defects.\cite{IwasakiT:PhysRevLett2017,LumannT:JPhysD2018,LonderoE:PhysRevB2018} In this context, both the Si- and Ge-vacancy defect have received increased interest.\cite{MersonTD:OptLett2013,IwasakiT:SciRep2015,PalyanovYN:SciRep2015,GreenBL:PhysRevLett2017, HausslerS:NewJPhys2017}\\
\indent Within condensed matter science, rare-earth (RE) elements receive a lot of attention as optical dopants, as they give rise to very sharp emission lines, for a wide variety of wavelengths from UV to IR.\cite{Jones:OptMater2006, SanaS:PhysRevB2009, OumaCNM:PhysBCondMat2014, CajzlJ:PhysChemChemPhys2017, CajzlJ:SurfInterAnal2018, TanX:AIPAdv2018} Furthermore, their filled $d$-shell shields the long lived excited states. In particular Eu and Gd are of great interest as both have a half-filled $4f$ shell. Luminescent Eu is used in inorganic scintillators, but also wide band gap semiconductors such as ZnO.\cite{ChaudhryA:PhysRevB2014, DharaS:Nanotech2014, VanpouckeDannyEP:2014f_Nanotechnology, BinnemansK:CoordChemRev2015,GuptaSK:RSCAdv2016, LorkeM:PhysRevB2016, YangY:OptMaterExpress2016} The high quantum efficiency and long photoluminescence decay times of RE elements (milliseconds, compared to nanoseconds for NV, SiV, or GeV) makes them very interesting color centres for quantum memory and repeater applications. Also in case of diamond, lanthanides are receiving attention as possible luminescent centres. Tan and co-workers calculated the zero-phonon line and emission wavelength of Pr-based defects in diamond using first principle methods.\cite{TanX:AIPAdv2018} The incorporation of Er and Yb was reported by Cajzl and co-workers.\cite{CajzlJ:PhysChemChemPhys2017, CajzlJ:SurfInterAnal2018} They showed Er to form multiple stable configurations in diamond and to present luminescence in the near infrared region.\cite{CajzlJ:PhysChemChemPhys2017} Ekimov \textit{et al.} investigated the catalytic properties of RE for the High Pressure High Temperature (HPHT) synthesis of diamond.\cite{Ekimov:MatterLett2017,EkimovEA:InorgMater2017} Recently, the incorporation of Eu in diamond during chemical vapour deposition (CVD) was reported by Magyar \textit{et al.} and Sedov \textit{et al.}\cite{MagyarA:NatComm2014, SedovV:DRM2017}\\
\indent Magyar and co-workers present the synthesis of Eu-doped diamond through the self assembly of a Eu containing precursor complex on the diamond surface, which is then overgrown with diamond during a standard CVD experiment.\cite{MagyarA:NatComm2014} The resulting samples show typical Eu fluorescence with a strong line at $612$ nm. Sedov and co-workers, on the other hand, incorporate \ce{EuF3} particles in diamond, while retaining the photoluminescent properties of Eu.\cite{SedovV:DRM2017} Ekimov and co-workers showed that Eu, Er and Tm exhibit catalytic activity during HPHT synthesis of diamond. However, in contrast to the previous two studies, they note that neither of these lanthanides shows photoluminescence, and they propose the photoluminescence observed earlier by Magyar and co-workers is not due to Eu but rather due to the \ce{Eu(DPA)3} complexes imbedded in the diamond samples. To this date, no detailed theoretical electronic structure calculations have been performed on Eu dopants in diamond.\\
\indent In this work, we present a combined theoretical--experimental study of Eu-doped diamond and to resolve the question whether Eu can form luminescent centres in diamond. RE elements have a complex electronic structure, with a $f$ electrons being shielded by the outermost $d$ electrons. Computational modeling can provide detailed insight in this electronic structure, with a clear correlation to the underlying atomic structure. We focus on the oxidation state of possible defects and the electronic structure in all its facets. Also the role of temperature on the defect stability is determined. %The full vibrational spectrum of the different systems is presented and decomposed to explicitly extract the dopant/defect-complex contributions.
We discuss several treatments for preparing diamond films, all aimed at achieving Eu-doping, and analyse the synthesized films by photoluminescence (PL) spectroscopy.\\
\begin{figure*}[t!]%
\includegraphics[width=15cm,keepaspectratio]{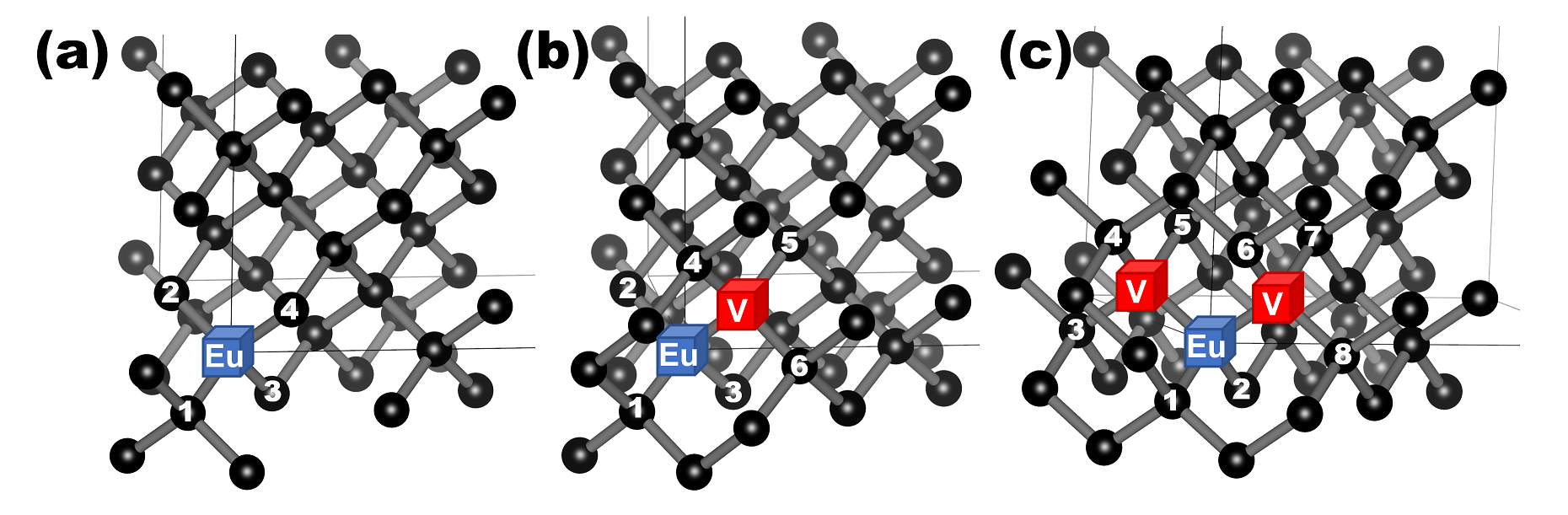}
\caption{(a) Ball-and-stick representation of the initial geometry of the (a) Eu-substitutional dopant, Eu$_{sub}$, (b) the Eu-vacancy defect, Eu$_{1V}$, and (c) the Eu-divacancy complex, Eu$_{2V}$. Red cubes indicate the position of the vacancies, blue cubes the position of the Eu atom, C atoms neighbouring the defect are numbered. The ball-and-stick images are created using VESTA.\cite{VESTA:JApplCryst2008}}
\label{fig:structure}
\end{figure*}

\section{Methodology}
\subsection{Theoretical model}
\indent In this work, we consider three models for Eu dopants in bulk diamond: (1) diamond with substitutional Eu doping, Eu$_{sub}$ , (2) diamond with an Eu-vacancy defect, Eu$_{1V}$, and (3) diamond with an Eu-divacancy complex, Eu$_{2V}$. Each of the three models are created in a conventional cubic $64$-atom cell by the substitution and removal of neighbouring C atoms. The initial local geometries of the three models is shown in Fig.~\ref{fig:structure}. For each of the models, a large number of possibly unpaired electrons are present. The Eu ion itself presents $7$ unpaired $f$ electrons, while the C atoms neighbouring the vacancies give rise to $3$ and $6$ dangling bonds for the single- and divacancy models, respectively. Since these unpaired electrons can give rise to many possible recombinations, various spin-configurations are considered as well. For each of the spin systems, the atomic geometry is optimised.\\
\subsection{Computational details}
\indent First-principles density functional theory (DFT) calculations are performed with the VASP code. We use the generalised gradient approximation (GGA) as constructed by Perdew, Burke and Ernzerhof (PBE) to describe the exchange and correlation behaviour of the valence electrons (the $2s^22p^2$ and $5s^25p^66s^24f^7$ electrons for C and Eu, respectively) during structure optimization and the calculation of the vibrational spectra.\cite{PBE_1996prl} Because the $f$ electrons of lanthanides are known to show strongly correlated behaviour, a Hubbard-U-type correction is included (DFT+U) for the Eu $f$ electrons in all PBE level calculations. We make use of the implementation suggested by Leichtenstein and co-workers,\cite{LiechtensteinAI:PhysRevB1995} using an on-site Coulomb parameter U=$7.397$ eV and an on-site exchange parameter J=$1.109$ eV for Eu, as suggested in the literature.\cite{JohannesMD:PhysRevB2005,LarsonP:JPhysCondMat2006, MagyarA:NatComm2014,YangY:OptMaterExpress2016} For C, no +U corrections are used during the final structure optimisation in the presented results, although the on-site Coulomb and exchange parameters, optimized for the C vacancy in diamond,\cite{VanpouckeDannyEP:2017d_DRM_DFTUVac} were used to stabilize the initial electronic structure optimisations. Because the choice of the U parameter has a significant influence on the electronic structure of a system,\cite{ChaudhryA:PhysRevB2014} we make use of the HSE06 hybrid functional for the calculation of high accuracy electronic structures, circumventing the issue of parametrisation.\cite{HeydJ:JChemPhys2003_HSE, Heyd:JChemPhys2005_HSE_BG} This range-separated hybrid functional is specifically designed for solids and has successfully been used for a wide range of materials, including defects in diamond.\cite{HendersonTM:PhysStatusSolidiB2011, GarzaA:JPhysChemLett2016, HendrickxVanpoucke:InorgChem2015, VanpouckeDannyEP:2017a_JPhysChemC, CzelejK:DRM2017, VanpouckeDannyEP:2017d_DRM_DFTUVac}\\
\indent The lattice parameters, cell volume, and ionic positions are optimized simultaneously using a conjugate gradient method with the energy criterion set to $1.0\times 10^{-7}$. As a result, the largest remaining force on a single atom is below $2$ meV/\AA.\cite{fn:forces} The vibrational spectrum is calculated using the finite difference method for constructing the hessian matrix. From the latter, the dynamical matrix at each point of first Brillouin zone is calculated using the in-house developed HIVE toolbox. The kinetic energy cutoff is set to $600$ eV, while the first Brillouin zone is sampled using a $4\times 4\times 4$ (structure optimisation and HSE06 calculations), $5\times 5\times 5$ (PBE+U static and vibrational calculations) or a $9\times 9\times 9$ (PBE+U density of states) Monkhorst-Pack special $k$-point grid. In case of the HSE06 calculations, no further reduction of the k-point mesh for the Hartree-Fock part is employed, as in our previous work.\cite{VanpouckeDannyEP:2017d_DRM_DFTUVac} Atomic charges are calculated using the Hirshfeld-I partitioning scheme.\cite{VanpouckeDannyEP:2013aJComputChem, VanpouckeDannyEP:2013bJComputChem, VanpouckeDannyEP:2019b_MicroMesoZeo} A charge convergence criterium of $1.0\times 10^{-5}$ electron is set, and charges are integrated on a logarithmic radial grid with atom-centred spherical shells of $1202$ grid-points.\cite{LebedevVI_grid:1999DokladyMath, BeckeAD:1987JCP}\\

\subsection{Experimental setup}
\begin{figure}[t!]%
\includegraphics[width=8cm,keepaspectratio]{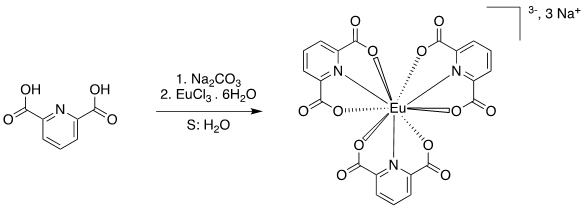}
\caption{Synthesis procedure of Eu(III) tris(dipicolinic acid) (\ce{Eu(DPA)3})}
\label{fig:EuDPA_synth}
\end{figure}

\indent Nanocrystalline diamond was chosen for the experimental study as the large surface area of the films, compared to small single crystal diamond substrates, allows for greater detection of the Eu fluorescence signature in PL. PL spectroscopy was performed in a Fluorolog 3-22 Tau system (Horiba Jobin-Yvon), with an excitation wavelength of $395$ nm, with the excitation and emission slit widths at $5.0$ and $0.5$s integration time. Follow up work investigating the incorporation of Eu in single crystal diamond, requiring more sensitive measurement techniques, is planned. Boron-doped nanocrystalline diamond (B:NCD) films are grown on \SI{1}{\centi\meter} $\times$ \SI{1}{\centi\meter} borosilicate glass substrates (Corning Eagle 2000). The substrates are cleaned using an RCA1 clean (5:1:1 - \ce{H2O}:\ce{NH3}:\ce{H2O2} at \SI{80}{\celsius} for $10$ min), flushed with milipore filtered deionized (DI) water, and then cleaned with an RCA2 clean (6:1:1 - \ce{H2O}:\ce{HCl}:\ce{H2O2} at \SI{80}{\celsius} for $10$ min). The substrates are then seeded with a colloidal suspension of detonation nanodiamonds, $6$--$7$ \si{\nano\meter} in size (NanoCarbon Institute Co., Ltd.) with a zeta potential of $(45 \pm 5)$ \si{\milli\volt}. The suspension is diluted in water to a concentration of \SI{0.33}{\gram\per\liter}. The seeding solution is drop-cast and then spun at $4000$ rpm on a commercially available spin coater (WS-400B-6NPP/LITE, Laurell Technologies Co.). During the first \SI{20}{\second} of spinning the sample is rinsed continuously with DI water, and then allowed to spin-dry.\\
\indent B:NCD films are deposited by microwave plasma assisted chemical vapor deposition (MPACVD) in an \nobreak{\textsc{ASTeX}} $6500$ series $2.45$ GHz microwave power supply reactor. Boron-doped diamond is used for these experiments to minimise the background luminescence from the negatively charged NV centre, which can be minimised through boron doping.\cite{Groot2014} Trimethylborane (TMB), dissolved in \ce{H2}, is used as the gas-phase boron-containing precursor. The vacuum chamber is pumped down prior to deposition by turbomolecular pump to a base pressure of $<$\SI{5e-6}{\milli\bar}. The forward power is set to \SI{3500}{\watt}, and the working pressure is held at $30$ Torr. The total flow rate of the predominantly \ce{H2} process gas is $500$ sccm, with a 1\% \ce{CH4} concentration and a TMB/\ce{CH4} ratio of $20,000$ ppm (2\%). The thickness of grown films is determined \emph{in situ} by observing the interference fringes of a \SI{405}{\nano\meter} laser, and the deposition is stopped after $47$ min at a thickness of \SI{150}{\nano\meter}. Samples grown in the same reactor under similar growth conditions\cite{Janssens2011,Drijkoningen2016} indicate that the B:NCD films have boron concentrations of approximately \SI{5e21}{\per\centi\meter\cubed}.\\
\indent Eu(III) tris(dipicolinic acid) (\ce{Eu(DPA)3}) is used as the Eu-containing precursor for all of the samples studied in this work. The \ce{Eu(DPA)3} complex is synthesized according to a slightly adapted procedure from the literature,\cite{Gaillard2013} as shown in Fig.~\ref{fig:EuDPA_synth}. Dipicolinic acid is deprotonated using an excess of sodium carbonate. After adding \ce{EuCl3} hexahydrate, the chlorine counterions are exchanged with three dipicolinic acid ligands, yielding the desired complex, as confirmed by \ce{^1H} NMR. Here we deliberately choose to employ sodium carbonate rather than sodium hydroxide to avoid formation of \ce{Eu(OH)3} as a side product.\\
\indent All samples are oxygen (O) terminated prior to Eu treatment in a Novascan PSD-UV Benchtop UV-ozone Cleaner for $1$ h. The electrostatic assembly method described in Magyar \textit{et al.}\cite{MagyarA:NatComm2014}, which makes use of an electronegative polymer material, is attempted on several samples. We use Poly(allylamine hydrochloride) (PAH) with average molecular weight $M_w = 17.5$ kDa (Sigma Aldrich), as $M_w = 50k$ PAH used in Magyar \textit{et al.} was not available at that time. The PAH is dissolved in water (0.1M by monomer) and the pH is adjusted to $8.0$ by the dropwise addition of ammonium hydroxide. \ce{Eu(DPA)3} is diluted to $0.1$M in water, and the pH adjusted to $3.8$ by dropwise addition of hydrochloric acid. The incubation steps in the PAH and \ce{Eu(DPA)3} are performed for $3$ h on one sample and $22$ h on another. We were, however, not able to measure the Eu signature by PL signal on samples treated this way before deposition, and so did not perform diamond deposition on the electrostatically assembled samples.\\
\indent The treatments investigated for samples exposed to the diamond growth plasma are summarised in  table~\ref{Table:samples}. \ce{Eu(DPA)3} is deposited by either drop-casting or spin-coating. For drop-casting, $0.2$ mL of the \ce{Eu(DPA)3} solution is applied to the O-terminated diamond surface, and allowed to evaporate while heating at \SI{100}{\celsius} on a hotplate in a cleanroom environment. The deposited \ce{Eu(DPA)3} film appears milky white in colour. For Sample (b) $0.2$ mL of the PAH solution is applied before the \ce{Eu(DPA)3} drop-casting procedure is performed. Spin-coating of \ce{Eu(DPA)3} is performed by applying $0.2$ mL to the sample surface and then spinning at $300$ rpm for $60$ s before spin drying at $4000$ rpm for $40$ s.\\
\indent The Eu-treated B:NCD films are then exposed to a diamond growth plasma for $10$ min under similar conditions as were employed in the work of Magyar \textit{et al}.\cite{MagyarA:NatComm2014} A plasma feedgas of $400$ sccm of \ce{H2} and $40$ sccm of \ce{CH4} is used, without the intentional addition of TMB, in the same diamond deposition reactor as was used to grow the B:NCD films. The forward microwave power is held at $2500$ W at a pressure of $17$ Torr.\\
\indent The substrate temperature during the initial substrate deposition was approximately \SI{700}{\celsius}, as estimated from measurements taken by infrared pyrometer (Cyclops $52$, Minolta, peak mode, $\varepsilon=0.6$), on similar samples during growth runs with the same deposition conditions, in the same reactor. During the $10$ min overgrowth step, a strong orange coloured optical emission, assumed to be due to the exposure of the Eu-containing precursor to the plasma environment, obscured the sample surface, preventing the assessment of the temperature by infrared emission during this phase.\\

\begin{table}[t!]
\caption{Summary of Eu-precursor deposition methods}
\begin{ruledtabular}
\begin{tabular}{@{}ccc@{}}
 & \multicolumn{2}{c}{Deposition Method}\\
Sample  & PAH & \ce{Eu(DPA)3} \\
\hline
(a) & N/A & Drop-cast \\
(b) & Drop-cast & Drop-cast \\
(c) & N/A & Spin-coated \\
\end{tabular}
\end{ruledtabular}
\label{Table:samples}
\end{table}

\begin{figure}[tb!]%
\includegraphics[width=8cm,keepaspectratio]{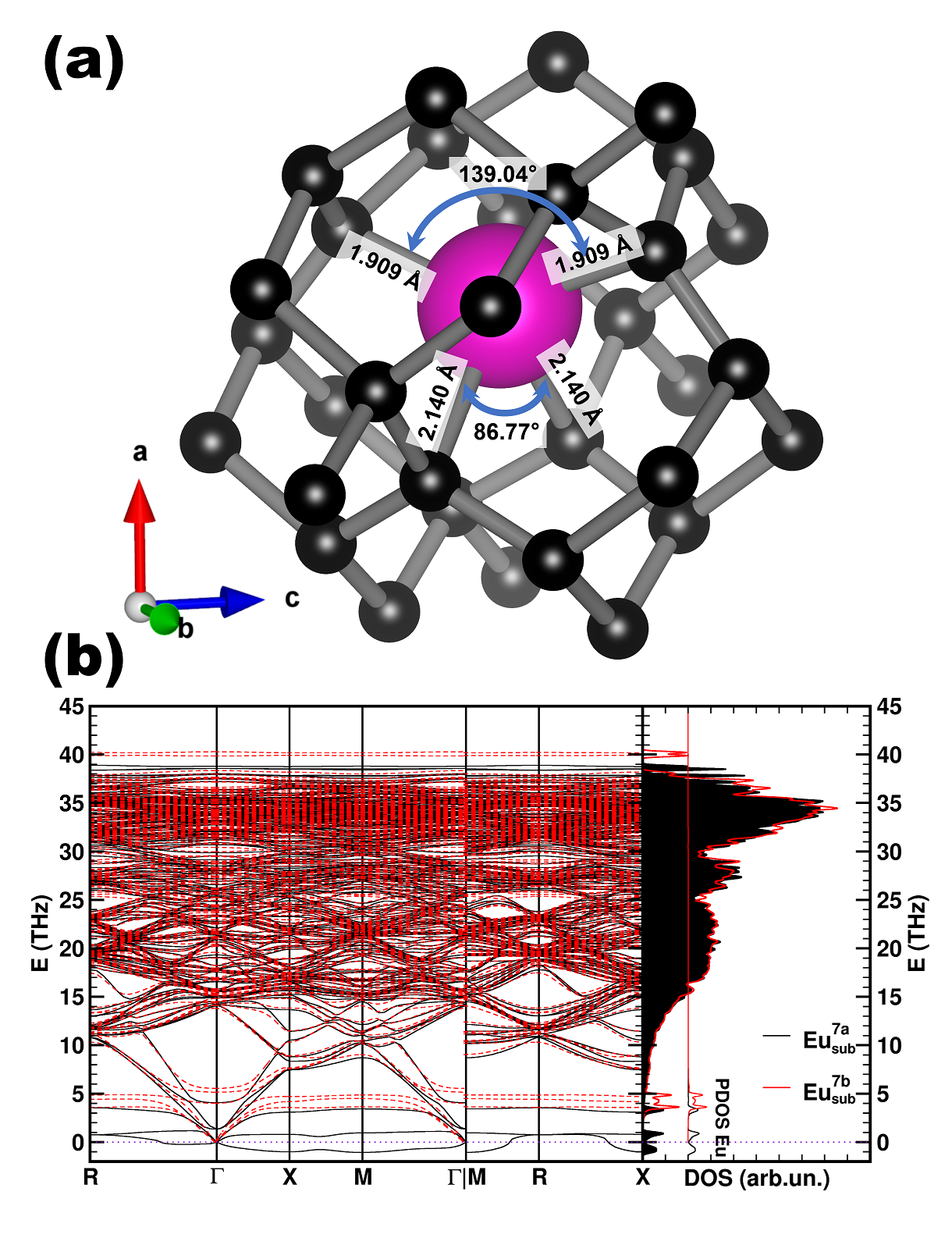}
\caption{(a) Ball and stick representation of the Eu$_{sub}^{7b}$ system, indicating the broken symmetry along the \textbf{a} lattice vector direction.
(b) Comparison of the phonon band structure and the phonon density of states of the Eu$_{sub}^{7a}$ (black) and Eu$_{sub}^{7b}$ (red) systems. The atom-projected density of states for the Eu ion is indicated, presenting a very distinct feature at low energies.}
\label{fig:VibSpectrEusub}
\end{figure}

\begin{table*}[tb!]
  \caption{Structural properties of the Eu defect in diamond: the coordination of the Eu ion (coord), the defect radius $R_d$ and volume $V_d$, the zero point energy (ZPE), formation energy ($E_f$) and strain energy ($E_s$).}\label{Table:properties}
  \begin{ruledtabular}
  \begin{tabular}{@{}l|rrrrrrrr@{}}
    %\hline
    & \multicolumn{3}{c}{} & \multicolumn{3}{c}{PBE+U} & \multicolumn{2}{c}{HSE06} \\
    & coord & $R_d$ & $V_d$   & $E_f$ & $E_s$ & ZPE  & $E_f$ & $E_s$ \\
    &       & \AA   & \AA$^3$ & (eV)  & (eV)  & (eV) & (eV)  & (eV) \\
    \hline
 Eu$_{sub}^{7a}$ & $4$ & $1.598$ & $24.177$ & $18.342$ & $12.738$ & $10.692$ & $19.400$ & $14.631$\\
 Eu$_{sub}^{7b}$ & $4$ & $1.504$ & $23.999$ & $18.161$ & $12.989$ & $10.721$ & $19.379$ & $14.901$\\
 Eu$_{sub}^9$ & $4$ & $1.598$ & $24.195$ & $18.457$ & $13.167$ & $10.686$ & $20.154$ & $15.100$\\
 Eu$_{1V}^7$ & $6$ & $1.368$ & $24.660$ & $13.264$ & $7.023$ & $10.630$ & $14.644$ & $9.082$\\
 Eu$_{1V}^9$ & $6$ & $1.366$ & $24.822$ & $13.700$ & $6.858$ & $10.620$ & $15.125$ & $8.573$\\
 Eu$_{2V}^{7p}$ & $6$ & $1.198$ & $27.244$ & $13.656$ & $7.465$ & $10.406$ & $15.179$ & $8.589$\\
 Eu$_{2V}^{7c}$ & $6$ & $1.208$ & $27.228$ & $13.704$ & $7.606$ & $10.394$ & $15.100$ & $8.780$\\
 Eu$_{2V}^9$  & $6$ & $1.244$ & $27.397$ & $13.764$ & $7.787$ & $10.374$ & $15.476$ & $9.127$\\
  \end{tabular}
  \end{ruledtabular}
\end{table*}

\begin{table*}[t!]
  \caption{Interatomic distance between Eu and its NN C atoms. The C atoms are indexed as in Fig.~\ref{fig:RelGeom_Mag}}\label{Table:bonds}
  \begin{ruledtabular}
  \begin{tabular}{@{}l|rrrrrrrr@{}}
    %\hline
    & $C_1$   & $C_2$ & $C_3$ & $C_4$  & $C_5$ & $C_6$ & $C_7$ & $C_8$ \\
    \hline
 Eu$_{sub}^{7a}$ & $2.050$ & $2.050$ & $1.982$ & $2.050$ &  &  &  &  \\
 Eu$_{sub}^{7b}$ & $2.140$ & $1.909$ & $1.909$ & $2.140$ &  &  &  &  \\
 Eu$_{sub}^9$    & $2.027$ & $2.027$ & $2.027$ & $2.120$ &  &  &  &  \\
 Eu$_{1V}^7$     & $2.194$ & $2.194$ & $2.194$ & $2.194$ & $2.194$ & $2.194$ &  &  \\
 Eu$_{1V}^9$     & $2.187$ & $2.187$ & $2.203$ & $2.203$ & $2.187$ & $2.187$ &  &  \\
 Eu$_{2V}^{7p}$  & $2.200$ & $2.200$ & $2.316$ & $2.316$ & $2.677$ & $2.314$ & $2.314$ & $2.670$  \\
 Eu$_{2V}^{7c}$  & $2.211$ & $2.211$ & $2.390$ & $2.389$ & $2.811$ & $2.274$ & $2.274$ & $2.540$  \\
 Eu$_{2V}^9$     & $2.247$ & $2.247$ & $2.308$ & $2.308$ & $2.683$ & $2.308$ & $2.308$ & $2.683$  \\
  \end{tabular}
  \end{ruledtabular}
\end{table*}
\section{Results and discussion}
\indent While modeling the defect(-complexes) at the quantum mechanical level, both atomic positions and electron distributions are taken into account. The latter aspect manifests itself in the spin-configuration of the system, which can have a significant impact on global materials properties in case $d$ or $f$ block elements are involved.\cite{VanpouckeDannyEP:2014e_Beilstein} For highly localized systems such as defects, initially unpaired electrons can form pairs to stabilize the system, as is the case for the C-vacancy in diamond.\cite{VanpouckeDannyEP:2017d_DRM_DFTUVac} The combination with a strongly correlated material, such as a transition metal or RE, can complicate the situation significantly. In case of the substitutional incorporation of Eu in diamond, there are $11$ electrons to consider ($7$ Eu $f$-electrons, and $4$ electrons in the bonds of the nearest neighbour C atoms). If one assumes Eu to be in the $2+$ oxidation state, then $2$ C electrons are left unpaired, giving rise to $9$ unpaired electrons. In case of Eu$^{3+}$, only $1$ C electron is left unpaired, as well as $6$ Eu $f$-electrons, giving a total of $7$ unpaired electrons. A similar mental exercise can be performed for the Eu-vacancy and Eu-divacancy models, leading to configurations with $7$, $9$, and $11$ unpaired electrons for the former, and $7$, $9$, $11$, and $13$ unpaired electrons for the latter. However, many of these spin configurations give rise to unphysical electronic structures.\cite{fn:unphys} For this reason, we only present the spin configurations with $7$ and $9$ unpaired electrons as these are present for each of the models. In this work, we indicate the number of unpaired electrons with a superscript in the system names.
\begin{table*}[tb!]
  \caption{Hirshfeld-I charges of the Eu defect in diamond in units of electron.\cite{VanpouckeDannyEP:2013aJComputChem, VanpouckeDannyEP:2013bJComputChem} The C atoms are indexed as in Fig.~\ref{fig:RelGeom_Mag}. The charge of the \emph{defect} is the sum of the Eu charge and the NN C ions ``bound'' to the Eu ion.}\label{Table:HICharges}
  \begin{ruledtabular}
  \begin{tabular}{@{}l|rrrrrrrrrr@{}}
    %\hline
    & Eu & defect & $C_1$   & $C_2$ & $C_3$ & $C_4$  & $C_5$ & $C_6$ & $C_7$ & $C_8$ \\
    \hline
 Eu$_{sub}^{7a}$ & $2.218$ & $-1.762$ & $-1.023$ & $-1.023$ & $-0.911$ & $-1.023$ &  &  &  &  \\
 Eu$_{sub}^{7b}$ & $2.228$ & $-1.484$ & $-1.085$ & $-0.771$ & $-0.771$ & $-1.085$ &  &  &  &  \\
 Eu$_{sub}^9$    & $2.148$ & $-1.626$ & $-0.906$ & $-0.906$ & $-0.906$ & $-1.056$ &  &  &  &  \\
 Eu$_{1V}^7$     & $2.294$ & $-2.318$ & $-0.768$ & $-0.768$ & $-0.770$ & $-0.770$ & $-0.768$ & $-0.768$ &  &  \\
 Eu$_{1V}^9$     & $2.286$ & $-2.302$ & $-0.728$ & $-0.728$ & $-0.838$ & $-0.838$ & $-0.728$ & $-0.728$ &  &  \\
 Eu$_{2V}^{7p}$  & $2.408$ & $-2.126$ & $-0.943$ & $-0.942$ & $-0.664$ & $-0.664$ & $-0.312$ & $-0.665$ & $-0.665$ & $-0.321$  \\
 Eu$_{2V}^{7c}$  & $2.148$ & $-1.825$ & $-0.795$ & $-0.792$ & $-0.570$ & $-0.572$ & $-0.202$ & $-0.622$ & $-0.622$ & $-0.468$  \\
 Eu$_{2V}^9$     & $2.100$ & $-1.764$ & $-0.686$ & $-0.686$ & $-0.623$ & $-0.623$ & $-0.330$ & $-0.623$ & $-0.623$ & $-0.331$  \\
  \end{tabular}
  \end{ruledtabular}
\end{table*}
\subsection{Structure and stability}
\indent In comparison to C, Eu is a very large atom,\cite{fn:atomsize} and as such it is expected to introduce a significant amount of strain into the system. To gain insight in this property, we calculate the defect size $R_d$,\cite{VanpouckeDannyEP:2012aApplSurfSci} the defect volume $V_d$, and the induced strain energy $E_s$. The defect volume is simply calculated as $V_d = V_{EuC} - n_C\cdot V_{C}$: the volume of the C atoms in the system subtracted from the system volume, using the volume of a single C atom in pure diamond as $V_{C}$. The strain energy is calculated by replacing the Eu atom with a single C atom, and calculating the difference in energy with the relaxed C configuration (\textit{i.e.}, in case of Eu$_{sub}$ the relaxed C configuration is pure diamond, while for the Eu$_{1V}$ and Eu$_{2V}$ this is the neutral C vacancy, and divacancy in diamond). The results are presented in table~\ref{Table:properties}.\\
\indent In the substitutional models (Eu$_{sub}^{7a}$ and Eu$_{sub}^{9}$), Eu remains $4$-coordinated, pushing the nearest neighbour (NN) C atoms slightly outward from their original position, resulting in a Eu-C bond length of about $2$ \AA(\textit{cf}. Table~\ref{Table:bonds}). Interestingly, the calculated size of the Eu ion is almost $1.6$ \AA, more than twice the size of a C atom. The bonds of the four NN C atoms with the next-NN C atoms reduces from the $1.55$ \AA\ in pristine diamond, to $1.50\pm0.01$ \AA, resulting in an almost planar geometry with the inter-C angles increasing to $118-120^{\circ}$. As is to be expected, the induced strain is very large, giving rise to an energy penalty of about $13$ ($15$) eV for PBE (HSE06) calculations.\\
\indent Interestingly, the Eu$_{sub}^{7a}$ system, although showing no imaginary vibrational frequencies at $\Gamma$ has a significant imaginary frequency throughout the rest of the first Brillouin zone (\textit{cf.},~Fig.~\ref{fig:VibSpectrEusub}). This is not the case for the Eu$_{sub}^{9}$ system, which shows no imaginary frequencies at all. Starting from the latter geometry we reoptimised the structure under the boundary condition of $7$ unpaired electrons to obtain the Eu$_{sub}^{7b}$ system, free of imaginary frequencies as well. As can be seen in table~\ref{Table:properties}, this system is more stable than the Eu$_{sub}^{7a}$ system, even though the strain energy increased. Furthermore, the local geometry of the Eu dopant has changed; Eu is still $4$-coordinated, but this time two C bonds are $1.9$ \AA\ in length, and two are $2.14$ \AA\ in length, giving rise to a different distortion of the tetragonal geometry. This difference between the Eu$_{sub}^{7a}$ and Eu$_{sub}^{7b}$ systems may be similar to the distortions predicted for Mn$^{2+}$ in GaAs.\cite{SubramanianH:JPhysCondMatter2013} Subramanian and Han note that the undistorted tetrahedral structure (the starting configuration for our Eu$_{sub}$ relaxations) is also an eigenstate of the total angular momentum operator, where its trifold degeneracy can be broken through a Jahn-Teller distortion, which we observe for the Eu$_{sub}^{7a}$ and Eu$_{sub}^{9}$ systems (\textit{cf.} Table~\ref{Table:bonds}). However, these authors note there exists a unidirectional compressive growth strain in the (Ga,Mn)As system, which results in a polarization gradient along the growth direction, and a splitting of the heavy and light hole subbands. It also leads to the As NN ``below'' and ``above'' the Mn ion to form different bonds. Considering the Eu$_{sub}^{7b}$ system, we find that, in contrast to Eu$_{sub}^{7a}$, the structure is not perfectly cubic. Relative to the Eu$_{sub}^{7a}$ system the system is slightly compressed along the \textbf{\emph{a}} lattice vector ($3.622$ vs. $3.629$ \AA), while slightly expanded along the \textbf{\emph{b}} and \textbf{\emph{c}} lattice vector ($3.632$ vs. $3.629$ \AA). Furthermore, the NN C ``below'' and ``above'' Eu, along the \textbf{\emph{a}} lattice vector, show the same symmetry breaking as is the case for the (Ga,Mn)As system, indicative of a strain-stabilized Jahn-Teller distortion (\textit{cf.} Table~\ref{Table:bonds}). This is corroborated by the strain energy presented in Table~\ref{Table:properties}, which is largest for the Eu$_{sub}^{7b}$ system.\\ %Finally, the trivalent nature of Eu, is also visible in the local deformation of the lattice. The 4-coordination is not perfectly tetragonal, but rather a compressed (stretched) tetragon for the Eu$_{sub}^{7}$ (Eu$_{sub}^{9}$) configuration.

\begin{figure}[tb!]%
\includegraphics[width=8cm,keepaspectratio]{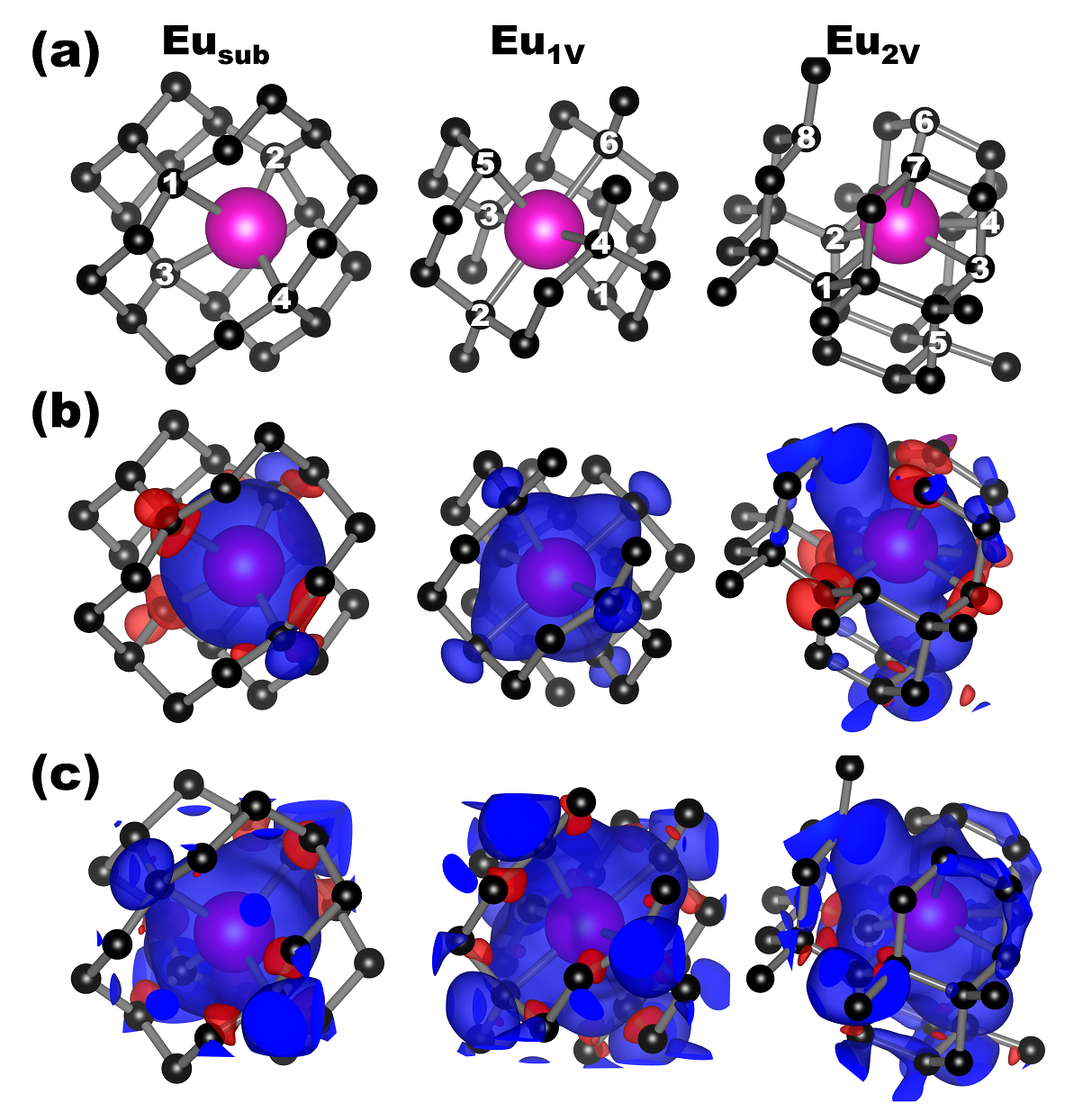}
\caption{(a)Ball-and-stick representation of the different models after structure optimisation. The NN C atoms are indicated. (b) The local magnetisation of the defect for Eu$_{sub}^{7b}$ (left), Eu$_{1V}^{7}$ (middle), and Eu$_{2V}^{7}$ (right). (c) The local magnetisation of the defect for Eu$_{sub}^{9}$ (left), Eu$_{1V}^{9}$ (middle), and Eu$_{2V}^{9}$ (right). Blue and red indicate the spin up and down components, respectively. All isosurfaces are taken at a value of 0.001 e.}
\label{fig:RelGeom_Mag}
\end{figure}

\indent In the Eu-vacancy model, the Eu ion shifts to the centre of the defect during relaxation, creating what is known as a split-vacancy defect (\textit{cf.} Fig.~\ref{fig:RelGeom_Mag}a). The Eu ion is 6-coordinated. This increase in coordination results in a reduced defect radius, albeit still larger than the Shannon crystal radius of Eu.\cite{Shannon:ACSA1976} The volume of the defect is slightly larger than for the substitutional case, showing the Eu ion just fills up all available space. The strain energy on the other hand reduces significantly, presenting a first indication that this configuration is preferred over the substitutional case. The Eu-C bond length increases to $2.19\pm0.01$ \AA, and the six NN C atoms are pushed outward from their initial position, resulting in a more planar configuration with the next NN C atoms. Each of the six NN C atoms presents two C-C bonds of $1.53$ \AA\ and one C-C bond of $1.48$ \AA, breaking the local trifold symmetry.\\

\indent The divacancy model presents a more complex reconstruction upon relaxation, as is shown in Fig.~\ref{fig:RelGeom_Mag}a. The different spin configurations give rise to very similar geometries. Similar to the Eu-vacancy model, the Eu ion moves toward the centre of the defect and fills out the available space as much as possible. However, although $8$ C atoms are available for bonding, only $6$ bonds are formed. Interestingly all $8$ C atoms neighbouring the defect are pushed outwards. As a result, the two C atoms without Eu-bond obtain an sp$^2$ configuration. The six Eu-C bonds are split into two groups: with two shorter bonds of $2.20$ \AA\ (Eu$_{2V}^{7p}$) and $2.25$ \AA\ (Eu$_{2V}^{9}$), and four longer bonds of $2.32$ \AA\ (Eu$_{2V}^{7p}$) and $2.31$ \AA\ (Eu$_{2V}^{9}$). In contrast, for the Eu$_{2V}^{7c}$ system the two shorter bonds are $2.21$ \AA, while the four longer bonds have varying lengths of $2.27$--$2.39$ \AA. This shows the impact of even slightly different spin configurations and electron distributions (\textit{cf.}~Table~\ref{Table:HICharges})\\
\indent In contrast to expectation, the strain energy of the divacancy system is roughly the same as for the single-vacancy system. This may originate from the fact that also in the divacancy model, Eu is only 6-coordinated, as well as the fact that the relaxed divacancy is significantly more stable than two next NN single vacancies. Furthermore, there are now $8$ C atoms with a significantly distorted local geometry, in contrast to only $6$ for the Eu-vacancy system. Taking the strain energy as well as the defect formation energy into consideration, we observe that the addition of more than one vacancy does not significantly improve the stability of the system, making the existence of these more complex systems less likely during CVD growth.\\
\begin{figure}[bt!]%
\includegraphics[width=8cm,keepaspectratio]{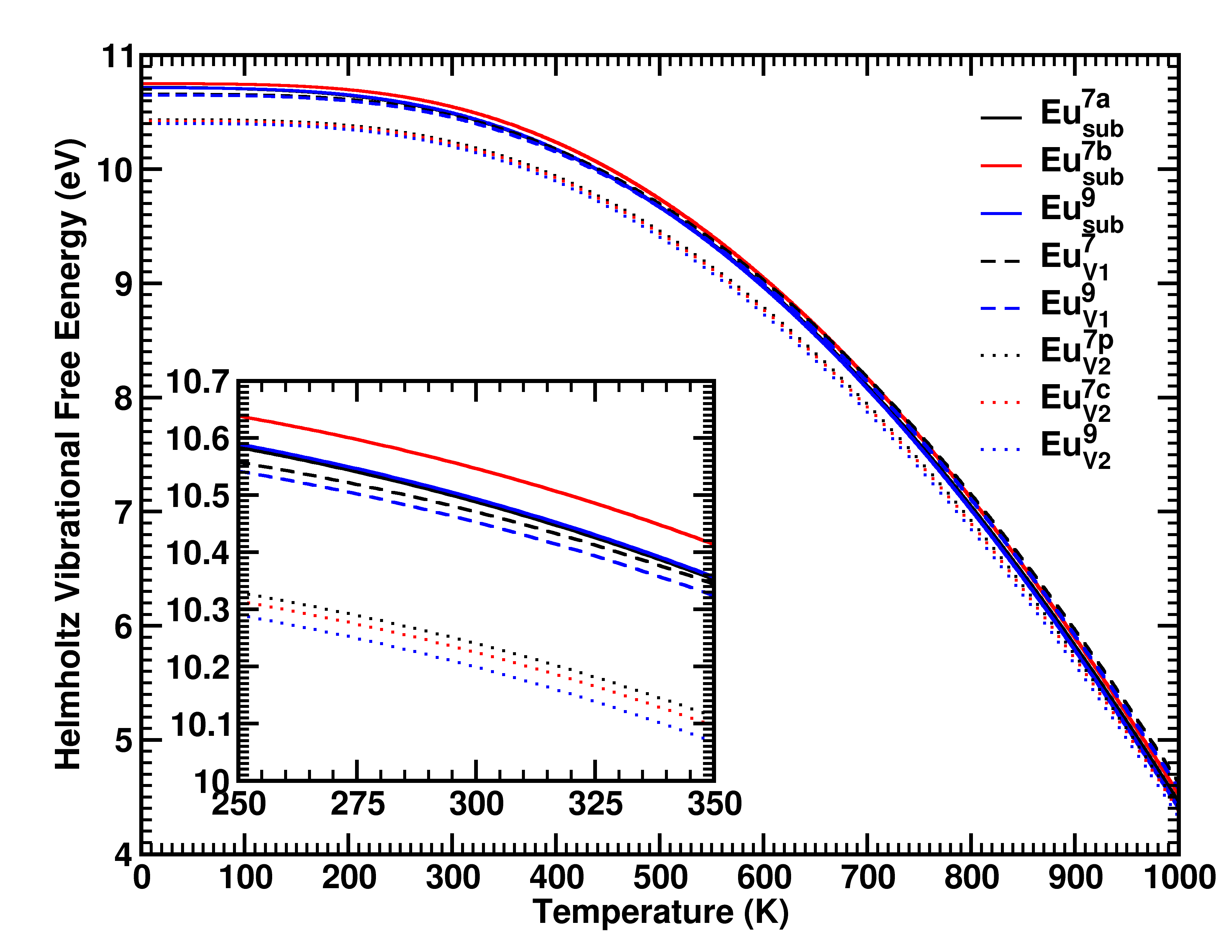}
\caption{The Helmholtz vibrational free energy as a function of the temperature, calculated based on the PBE vibrational spectrum of the full first Brillouin zone. A zoomed in region around room temperature is shown in the inset.}
\label{fig:EvibHelm}
\end{figure}
\indent The temperature contributions to the total energy are estimated as the Helmholtz vibrational free energy and are shown in Fig.~\ref{fig:EvibHelm}. The Helmholtz vibrational free energy contributions of all systems show the same qualitative temperature dependence. The inclusion of vacancies lowers the Helmholtz free energy at low temperatures. Note that this lowering originates with the internal vibrational energy, as the vibrational entropy term is roughly the same for all systems. The divacancy systems have the lowest Helmholtz vibrational free energy over the entire range presented. At low temperatures the single vacancy systems have a lower Helmholtz vibrational free energy than the substitutional systems. However, at temperatures above 400K this order of the Helmholtz vibrational free energy reverses. As can be seen in the inset, the energy differences are much smaller than those present in the formation energies. As such, we do not expect temperature effects to play an important role in the overall stability order of the different Eu-defect models presented in this work.\\
\begin{figure}[tb!]%
\includegraphics[width=8cm,keepaspectratio]{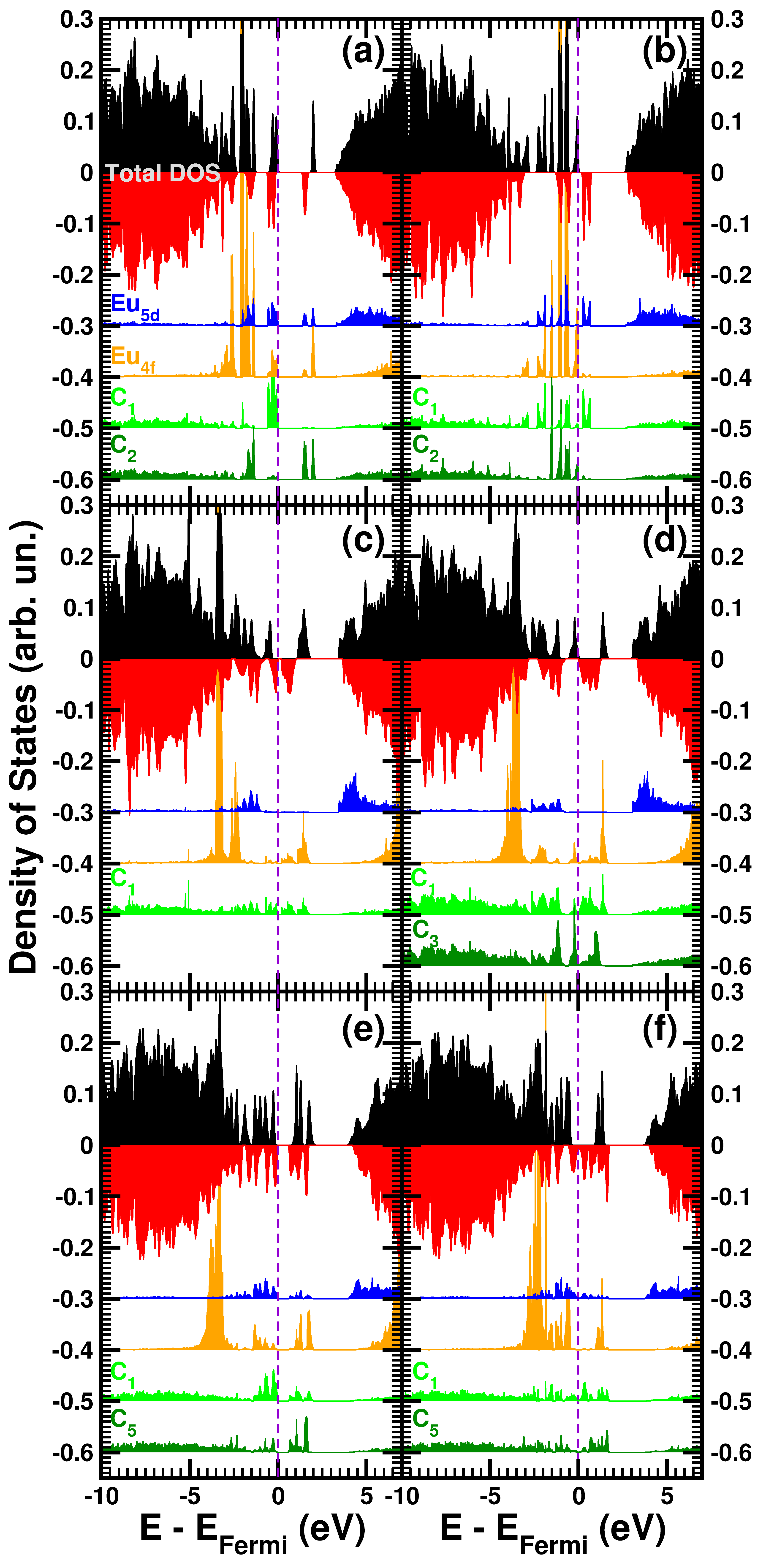}
\caption{The spin polarized total density of states of the Eu-doped diamond systems ((a) Eu$_{sub}^{7b}$, (b) Eu$_{sub}^{9}$, (c) Eu$_{1V}^{7}$, (d)  Eu$_{1V}^{9}$, (e)  Eu$_{2V}^{7c}$, and (f)  Eu$_{2V}^{9}$ ), as calculated using the HSE06 functional. The projected density of states (PDOS) is also shown for the Eu $5d$-states (blue) and the Eu $4f$-states (orange). In addition, the total PDOS for selected NN C ions is shown in green. The Fermi level is defined as the top of the defect valence band.}
\label{fig:DOSHSE}
\end{figure}

\begin{table}[bt!]
  \caption{Band gap size of the Eu defect in diamond, using the HSE06 hybrid functional. [units: eV]}\label{Table:bandgap}
  \begin{ruledtabular}
  \begin{tabular}{@{}l|rrr@{}}
    %\hline
    & spin & spin & effective \\
    & up   & down & band gap \\
    \hline
 Eu$_{sub}^{7a}$ & $1.620$ & $1.350$ & $1.300$ \\
 Eu$_{sub}^{7b}$ & $1.940$ & $1.550$ & $1.400$ \\
 Eu$_{sub}^9$    & $2.690$ & $0.710$ & $0.250$ \\
 Eu$_{1V}^7$     & $1.490$ & $0.240$ & $0.240$ \\
 Eu$_{1V}^9$     & $1.220$ & $0.740$ & $0.000$ \\
 Eu$_{2V}^{7p}$  & $0.631$ & $0.951$ & $0.531$ \\
 Eu$_{2V}^{7c}$  & $0.991$ & $0.591$ & $0.591$ \\
 Eu$_{2V}^9$     & $1.381$ & $0.000$ & $0.000$ \\
  \end{tabular}
  \end{ruledtabular}
\end{table}
\subsection{Electronic structure}
\indent For Eu to act as a luminescent centre in the visible region of the spectrum in diamond,\cite{fn:visible} it should be in the ${3+}$ oxidation state. Furthermore, the Eu $4f$ and $5d$-states are expected to be located in the band gap. To investigate the first aspect, we make use of the Hirshfeld-I atoms-in-molecules (AIM) partitioning method.\cite{VanpouckeDannyEP:2013aJComputChem, VanpouckeDannyEP:2013bJComputChem, VanpouckeDannyEP:2019b_MicroMesoZeo} This is a robust AIM method which provides atomic charges, larger than those obtained with the Mulliken or normal Hirshfeld methods, but still smaller than the formal charge.\cite{VanpouckeDannyEP:2019b_MicroMesoZeo} As such we can use this method to infer the oxidation state of the Eu ion.\\
\indent In Table~\ref{Table:HICharges}, we present the Hirshfeld-I charges of the Eu ion and its NN C atoms, as well as the total charge of the complex (Eu+NN C). For each of the considered systems, the charge of the Eu ion exceeds $+2$ indicating Eu to be in the Eu$^{3+}$ oxidation state. All NN C atoms on the other hand have a large negative charge, creating a dipole at the defect site. Interestingly, these negative charges are much larger than would be expected if only charge transfer from the Eu ion is involved. As a result, the entire defect(-complex) has a significant global negative charge (\textit{cf.}~Table~\ref{Table:HICharges}). If all NN C atoms are considered, the defect charge steadily decreases with the defect size as more negatively charged C are considered. However, if only the NN C ions bound to the Eu ion are considered, then the total defect charge has a minimum for the Eu-vacancy systems. In this work, we only consider single defects, due to the relatively small super cell used. As a result, all the additional negative charge accumulated on the Eu-defect complex has been drawn from the nearby C atoms. However, in an experimental sample, also other defect-centres will be present, such as N or P which may provide the additional charge. However, a detailed study of such interactions is beyond the scope of the current work, and will be taken up in future work.\\
\indent The total density of states (DOS) and projected DOS (PDOS) of the different systems are shown in Fig.~\ref{fig:DOSHSE}.\cite{fn:DOS} It is immediately clear that for the systems with $9$ unpaired electrons there is no or only a very small effective band gap present (\textit{cf.} Table~\ref{Table:bandgap}). With the exception of the divacancy system (which appears to be a half-metal), this is due to the relative alignment of the majority (up) and minority (down) spin densities. The substitutional system with $7$ unpaired electrons shows the largest effective band gap, as well as the largest individual spin band gaps of about $1.5$--$2.0$ eV. At this point, it is important to note that the hypersensitive transition of Eu$^{3+}$ luminescence obeys the selection rules $|\Delta S|= 0$, $|\Delta L|\leq 2$, and $|\Delta J|\leq 2$.\cite{BinnemansK:CoordChemRev2015} As a straightforward consequence, this means that it are the spin band gaps which are relevant for possible luminescence. Therefore the spin band gaps of the Eu$^{7}_{sub}$ systems are perfectly in the energy range of the photoluminescence observed for Eu$^{3+}$. Furthermore, the defect states seen in the band gap (\textit{cf.},~Fig.~\ref{fig:DOSHSE}a) have a Eu $5d$ and $4f$ character, again what should be expected in the presence of luminescent Eu.\\
\indent In contrast, the gap states for the Eu-split-vacancy system show no $5d$ character, but a combination of mainly Eu $5p$ (not shown) and some $4f$ character, as well as C $2p$ character, indicative of the Eu-C bonding. The most pronounced Eu $5d$ and $4f$ contributions, relevant for Eu$^{3+}$ luminescence in the Eu-split-vacancy model, are found at the valence ($4f$ states) and conduction ($5d$ states) band edges of the host diamond band gap, resulting in a large separation between the $d$ and $f$ states. As such, it is unlikely, that this system will present the luminescence expected around $611$ nm of Eu$^{3+}$. \\
\indent In case of the Eu-divacancy system, the gap states again show a clear Eu $5d$ and $4f$ character, as well as a strong C $2p$ character showing these states to be involved in the bonding of the Eu ion. In case of the Eu$_{2V}^{9}$ system, the small band gap in the majority channel originates from the Eu $4f$ states, while the half-metal behaviour of the minority channel originates from the Eu $5d$ states. In contrast, the Eu$_{2V}^{7p}$ and Eu$_{2V}^{7c}$ systems are semiconducting for both spin channels. It is also interesting to note that the band gap for both spin channels is determined by the Eu $d$ states. Unfortunately, the spin band gaps of the Eu$_{2V}^{7p}$ and Eu$_{2V}^{7c}$ systems ($0.6$ to $1.0$ eV) are too small to allow for the expected Eu$^{3+}$ luminescence at $611$--$614$ nm, which requires a band gap of about $2.02$ eV. In contrast, these smaller band gaps could be a hint toward the potential for IR applications of this system. As we did not perform PL measurements in this region, further corroboration of this theoretical result is still needed.\\
\indent If we look back at the local geometry of the Eu-divacancy complexes it is hard not to notice the large amount of sp$^2$-like C present. In a polycrystalline diamond film, grain boundaries provide naturally occurring regions with a high density of sp$^2$ C. As such, one could speculate that such a more amorphous or diamond-like carbon (DLC) region could provide the required environment for Eu to form a luminescent centre. This will be the subject of future work.\\
\begin{figure}[bt!]%
\includegraphics[width=8cm,keepaspectratio]{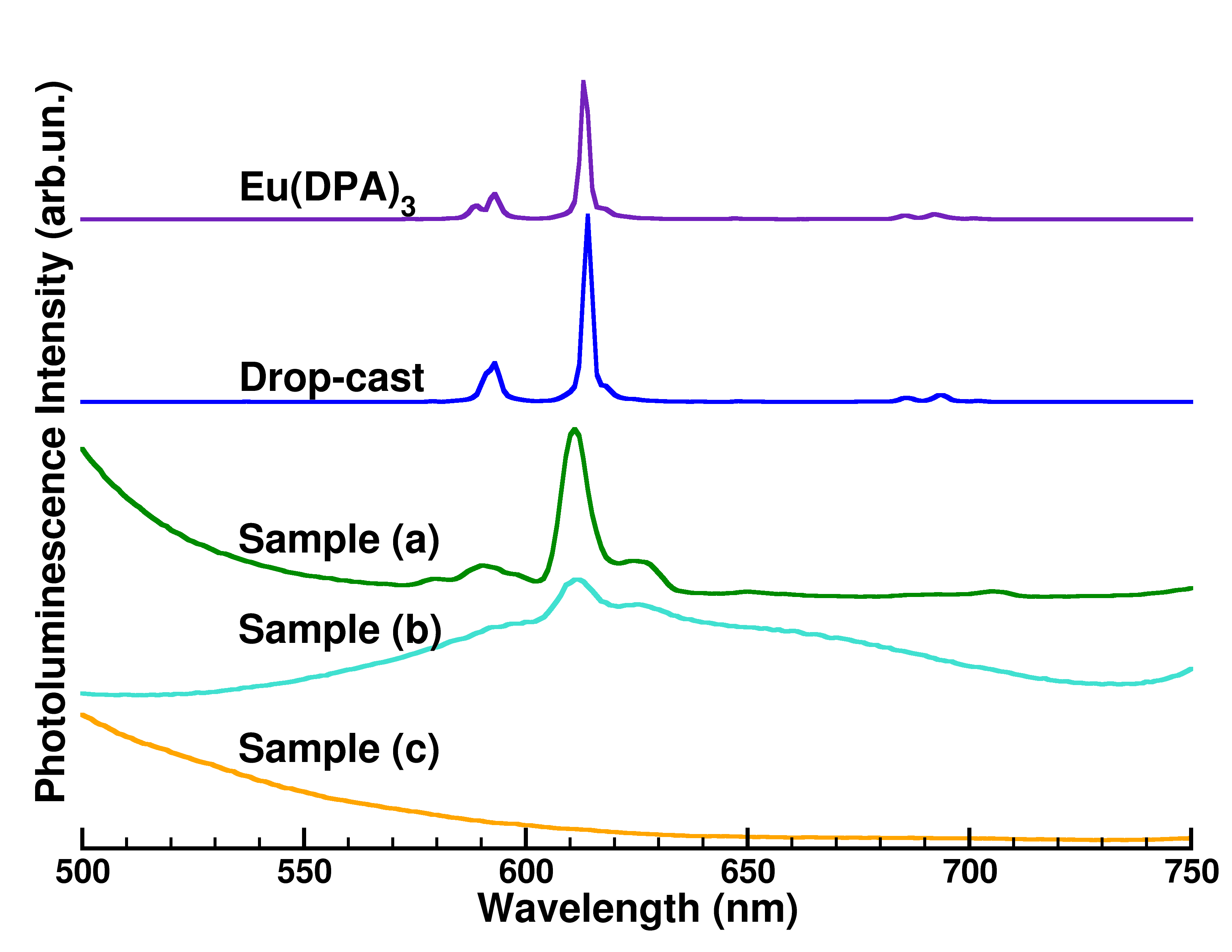}
\caption{Photoluminescence spectroscopy of the \ce{Eu(DPA)3} precursor solution, a sample treated with the \ce{Eu(PDA)3} by drop-casting but not exposed to the diamond deposition plasma, and Samples (a), (b) and (c).}
\label{fig:PL}
\end{figure}
\subsection{Experimental results}
\indent Photoluminescence spectroscopy was performed on the \ce{Eu(DPA)3} solution, as shown in Fig.~\ref{fig:PL}. The \ce{Eu(DPA)3} precursor shows the characteristic peaks for Eu including the main peak at approximately $614$ nm, as previously been reported in the literature.\cite{MagyarA:NatComm2014} The PL of a diamond substrate which was not grown, but was treated by drop-casting also shows the same PL signature once the precursor is deposited on the surface. PL from the three samples described in  table~\ref{Table:samples} is also shown in Fig.~\ref{fig:PL}. Sample (a), which was drop-cast, shows the strongest Eu signature, at $611$ nm, indicating that some Eu-containing material remained after the diamond deposition process. Sample (b) also shows a peak at approximately $611$ nm, indicating the presence of Eu, however the background PL signal is much higher in this sample. We attribute this to the co-deposition with PAH in this sample. This indicates that the PAH is not necessary for the detection of an Eu PL signature following exposure to a growth plasma. Moreover, the use of PAH may decrease the PL signal relative to the background. Any Eu signature for Sample (c) was below the detection threshold. This could signify the incorporation of Eu as Eu-split-vacancy or Eu-divacancy defect, as expected from our calculations. Alternately, the presence of the Eu luminescence peak does not indicate that the Eu is incorporated in the diamond for both Samples (a) and (b).\cite{EkimovEA:InorgMater2017} However, the PL spectrum does change relative to the signature prior to growth. The main peak position shifts slightly toward lower wavelengths, from $614$ to $611$ nm, and a more pronounced shoulder of the peak, at about $625$ nm, appears. These changes are not attributable to a change in calibration, as the same calibration was used on the same day for the drop-cast sample and Samples (a)-(c). We note that a shoulder peak at approximately $625$ nm can also been seen in the PL spectra of Eu-treated diamond after exposure to a growth plasma in the work of Magyar \textit{et al}.\cite{MagyarA:NatComm2014} The presence of B co-doping, added to passivate SiV and NV defects of these samples, is expected to have a limited impact on the PL of Eu, as the luminescence of Eu is inherent to the electronic structure of the Eu ion (\textit{i.e.}, transition between Eu $d$  and $f$ states). As such, although the incorporated B may give rise to additional states in the band gap, modifying the global electronic structure, these states are not directly involved in the photoluminescence of the $d-f$ transition of Eu.\\

\section{Conclusion}
\indent In this work, we present a combined theoretical--experimental study of Eu-doped diamond. We show that the high defect formation energy is mainly due to the induced defect strain energy. Interestingly, the strain energy and defect formation energy show no further improvement beyond the incorporation of the first vacancy (\textit{i.e.}, the Eu-vacancy). Inclusion of temperature effects through the vibrational energy does not change the qualitative stability picture of the different configurations and defect models. In future work, the role of the defect concentration on the stability will be considered.\\
\indent In the case of substitutionally doped diamond, Eu is $4$-coordinated despite the large induced strain. Furthermore, a strain-stabilized Jahn-Teller distortion gives rise to a polarisation of the Eu defect, similar to what is observed in the (Ga,Mn)As system. The Eu-vacancy and Eu-divacancy defects on the other hand both give rise to $6$-coordinated Eu, even though two more C are accessible for bonding in the latter case. Electron density partitioning shows Eu to be positively charged and in the $3+$ oxidation state. The nearest neighbour shell of C atoms on the other hand is negatively charged, leading to a strong local polarization. In addition, the full defect-complex is negatively charged, with a charge in the range of $-1.5$ to $-2.3$ electron.\\
%\indent The electronic band structure of the different defect models was compared, showing that substitutional doping with Eu is the most promising candidate for the observation of the $611$ nm luminescence of Eu$^{3+}$. The Eu-vacancy model on the other hand is not expected to present any luminescence related to the typical Eu $d-f$ transitions, while the Eu-divacany system could again give rise to luminescent transitions, albeit at a much longer wavelengths. Based on the Eu-divacancy results, we suspect the incorporation of Eu in the grain boundaries may give rise to the formation of luminescent centres, a topic for future investigations.\\
\indent Even though the Eu ion is shown to be in the $3+$ oxidation state in each model, the electronic band structure presents a rather complex picture. As a result, the answer to whether Eu forms a luminescent centre is not unique, and depends strongly on the local geometry and spin configuration. In summary:
\begin{itemize}
  \item Substitutional doping with Eu is the most promising candidate for the observation of the $611$ nm luminescence of Eu$^{3+}$.
  \item The Eu-vacancy model is not expected to present any luminescence related to the typical Eu $d-f$ transitions.
  \item The Eu-divacany system could give rise to luminescent transitions, albeit at a much longer wavelengths (\textit{i.e.}, IR).
\end{itemize}
Based on the Eu-divacancy results, we suspect that incorporation of Eu in the grain boundaries or in diamond-like carbon may give rise to the formation of luminescent centres due to the presence of sp$^2$ carbon, a topic for future investigations.\\
\indent Diamond films are prepared with the \ce{Eu(DPA)3} precursor using three different methods and then exposed to a diamond growth plasma for $10$ min. PL measurements indicate that Eu-containing material remains in samples treated with drop-casting. Changes in the PL spectrum after exposure to the growth plasma provide preliminary evidence for the potential incorporation of Eu in the diamond film, however additional experimental work is needed to fully prove Eu inclusion in the diamond lattice. Future work examining and optimising the incorporation of Eu in single crystal diamond is also planned. Treatment with PAH is found not to be required to demonstrate the Eu signature, and increased the background PL when drop-cast under the \ce{Eu(DPA)3} precursor.\\

\section*{Prime novelty statement}
\indent Eu doping in diamond has been studied by means of high accuracy first-principles calculations. The local defect geometry is crucial with regard to the possible photoluminescence of Eu, even though Eu is shown to present a $3+$ oxidation state for all defect models. The incorporation of Eu in diamond was also studied experimentally, indicating that Eu containing material can remain after exposure to diamond growth plasma if drop-cast on the surface. Furthermore, the presence of PAH is not required, and increases the PL background.

%Special thanks etc
%----------------------------------------------------------------------
\begin{acknowledgements}
\indent This work was financially supported by the Hasselt University Special Research Fund (BOF) and the Methusalem NANO network. SSN is a Newton International Fellow of the Royal Society. JR is an FWO PhD fellow (n$^{\circ}$1S13716N). The computational resources and services used in this work were provided by the VSC (Flemish Supercomputer Center), funded by the Research Foundation - Flanders (FWO) and the Flemish Government -- department EWI.
\end{acknowledgements}
%----------------------------------------------------------------------

%*************************************************************************************************************
% BIBLIOGRAPHY
%*************************************************************************************************************
% Put in \nocite{*} so all entries in the bibliography are included
%\nocite{*}
% This GATHER command is useful for when you want to use WinEdt's Gather functionality, i.e., type
% \cite{} and a popup box appears with all of your citations to choose from.  Leave the % on the next line.
% The commented way of writing Gather is the only correct way of doing it !!!
%GATHER{danny.bib}
%GATHER{notesEuDoping.bib}
%GATHER{shannon.bib}
%%\bibliographystyle{pss}
%%\bibliographystyle{ieeetr}
\bibliography{danny,notesEuDoping,shannon}

\begin{thebibliography}{65}%
\makeatletter
\providecommand \@ifxundefined [1]{%
 \@ifx{#1\undefined}
}%
\providecommand \@ifnum [1]{%
 \ifnum #1\expandafter \@firstoftwo
 \else \expandafter \@secondoftwo
 \fi
}%
\providecommand \@ifx [1]{%
 \ifx #1\expandafter \@firstoftwo
 \else \expandafter \@secondoftwo
 \fi
}%
\providecommand \natexlab [1]{#1}%
\providecommand \enquote  [1]{``#1''}%
\providecommand \bibnamefont  [1]{#1}%
\providecommand \bibfnamefont [1]{#1}%
\providecommand \citenamefont [1]{#1}%
\providecommand \href@noop [0]{\@secondoftwo}%
\providecommand \href [0]{\begingroup \@sanitize@url \@href}%
\providecommand \@href[1]{\@@startlink{#1}\@@href}%
\providecommand \@@href[1]{\endgroup#1\@@endlink}%
\providecommand \@sanitize@url [0]{\catcode `\\12\catcode `\$12\catcode
  `\&12\catcode `\#12\catcode `\^12\catcode `\_12\catcode `\%12\relax}%
\providecommand \@@startlink[1]{}%
\providecommand \@@endlink[0]{}%
\providecommand \url  [0]{\begingroup\@sanitize@url \@url }%
\providecommand \@url [1]{\endgroup\@href {#1}{\urlprefix }}%
\providecommand \urlprefix  [0]{URL }%
\providecommand \Eprint [0]{\href }%
\providecommand \doibase [0]{http://dx.doi.org/}%
\providecommand \selectlanguage [0]{\@gobble}%
\providecommand \bibinfo  [0]{\@secondoftwo}%
\providecommand \bibfield  [0]{\@secondoftwo}%
\providecommand \translation [1]{[#1]}%
\providecommand \BibitemOpen [0]{}%
\providecommand \bibitemStop [0]{}%
\providecommand \bibitemNoStop [0]{.\EOS\space}%
\providecommand \EOS [0]{\spacefactor3000\relax}%
\providecommand \BibitemShut  [1]{\csname bibitem#1\endcsname}%
\let\auto@bib@innerbib\@empty
%</preamble>
\bibitem [{\citenamefont {Childress}\ and\ \citenamefont
  {Hanson}(2013)}]{ChildressH:MRSBull2013}%
  \BibitemOpen
  \bibfield  {author} {\bibinfo {author} {\bibfnamefont {Lilian}\ \bibnamefont
  {Childress}}\ and\ \bibinfo {author} {\bibfnamefont {Ronald}\ \bibnamefont
  {Hanson}},\ }\bibfield  {title} {\enquote {\bibinfo {title} {Diamond {NV}
  centers for quantum computing and quantum networks},}\ }\href {\doibase
  10.1557/mrs.2013.20} {\bibfield  {journal} {\bibinfo  {journal} {MRS
  Bulletin}\ }\textbf {\bibinfo {volume} {38}},\ \bibinfo {pages} {134–138}
  (\bibinfo {year} {2013})}\BibitemShut {NoStop}%
\bibitem [{\citenamefont {Bougas}\ \emph {et~al.}(2018)\citenamefont {Bougas},
  \citenamefont {Wilzewski}, \citenamefont {Dumeige}, \citenamefont {Antypas},
  \citenamefont {Wu}, \citenamefont {Wickenbrock}, \citenamefont {Bourgeois},
  \citenamefont {Nesladek}, \citenamefont {Clevenson}, \citenamefont {Braje},
  \citenamefont {Englund},\ and\ \citenamefont
  {Budker}}]{BougasL:Micromachines2018}%
  \BibitemOpen
  \bibfield  {author} {\bibinfo {author} {\bibfnamefont {Lykourgos}\
  \bibnamefont {Bougas}}, \bibinfo {author} {\bibfnamefont {Alexander}\
  \bibnamefont {Wilzewski}}, \bibinfo {author} {\bibfnamefont {Yannick}\
  \bibnamefont {Dumeige}}, \bibinfo {author} {\bibfnamefont {Dionysios}\
  \bibnamefont {Antypas}}, \bibinfo {author} {\bibfnamefont {Teng}\
  \bibnamefont {Wu}}, \bibinfo {author} {\bibfnamefont {Arne}\ \bibnamefont
  {Wickenbrock}}, \bibinfo {author} {\bibfnamefont {Emilie}\ \bibnamefont
  {Bourgeois}}, \bibinfo {author} {\bibfnamefont {Milos}\ \bibnamefont
  {Nesladek}}, \bibinfo {author} {\bibfnamefont {Hannah}\ \bibnamefont
  {Clevenson}}, \bibinfo {author} {\bibfnamefont {Danielle}\ \bibnamefont
  {Braje}}, \bibinfo {author} {\bibfnamefont {Dirk}\ \bibnamefont {Englund}}, \
  and\ \bibinfo {author} {\bibfnamefont {Dmitry}\ \bibnamefont {Budker}},\
  }\bibfield  {title} {\enquote {\bibinfo {title} {On the {P}ossibility of
  {M}iniature {Diamond-Based Magnetometers} {U}sing {W}aveguide
  {G}eometries},}\ }\href {http://www.mdpi.com/2072-666X/9/6/276} {\bibfield
  {journal} {\bibinfo  {journal} {Micromachines}\ }\textbf {\bibinfo {volume}
  {9}},\ \bibinfo {pages} {276} (\bibinfo {year} {2018})}\BibitemShut {NoStop}%
\bibitem [{\citenamefont {Aharonovich}\ and\ \citenamefont
  {Neu}(2014)}]{AharonovichI:AdvOptMater2014}%
  \BibitemOpen
  \bibfield  {author} {\bibinfo {author} {\bibfnamefont {Igor}\ \bibnamefont
  {Aharonovich}}\ and\ \bibinfo {author} {\bibfnamefont {Elke}\ \bibnamefont
  {Neu}},\ }\bibfield  {title} {\enquote {\bibinfo {title} {Diamond
  {N}anophotonics},}\ }\href {\doibase 10.1002/adom.201400189} {\bibfield
  {journal} {\bibinfo  {journal} {Adv. Opt. Mater.}\ }\textbf {\bibinfo
  {volume} {2}},\ \bibinfo {pages} {911--928} (\bibinfo {year}
  {2014})}\BibitemShut {NoStop}%
\bibitem [{\citenamefont {Pfaff}\ \emph {et~al.}(2014)\citenamefont {Pfaff},
  \citenamefont {Hensen}, \citenamefont {Bernien}, \citenamefont {van Dam},
  \citenamefont {Blok}, \citenamefont {Taminiau}, \citenamefont {Tiggelman},
  \citenamefont {Schouten}, \citenamefont {Markham}, \citenamefont {Twitchen},\
  and\ \citenamefont {Hanson}}]{PfaffW:Science2014}%
  \BibitemOpen
  \bibfield  {author} {\bibinfo {author} {\bibfnamefont {W.}~\bibnamefont
  {Pfaff}}, \bibinfo {author} {\bibfnamefont {B.~J.}\ \bibnamefont {Hensen}},
  \bibinfo {author} {\bibfnamefont {H.}~\bibnamefont {Bernien}}, \bibinfo
  {author} {\bibfnamefont {S.~B.}\ \bibnamefont {van Dam}}, \bibinfo {author}
  {\bibfnamefont {M.~S.}\ \bibnamefont {Blok}}, \bibinfo {author}
  {\bibfnamefont {T.~H.}\ \bibnamefont {Taminiau}}, \bibinfo {author}
  {\bibfnamefont {M.~J.}\ \bibnamefont {Tiggelman}}, \bibinfo {author}
  {\bibfnamefont {R.~N.}\ \bibnamefont {Schouten}}, \bibinfo {author}
  {\bibfnamefont {M.}~\bibnamefont {Markham}}, \bibinfo {author} {\bibfnamefont
  {D.~J.}\ \bibnamefont {Twitchen}}, \ and\ \bibinfo {author} {\bibfnamefont
  {R.}~\bibnamefont {Hanson}},\ }\bibfield  {title} {\enquote {\bibinfo {title}
  {Unconditional quantum teleportation between distant solid-state quantum
  bits},}\ }\href {\doibase 10.1126/science.1253512} {\bibfield  {journal}
  {\bibinfo  {journal} {Science}\ }\textbf {\bibinfo {volume} {345}},\ \bibinfo
  {pages} {532--535} (\bibinfo {year} {2014})}\BibitemShut {NoStop}%
\bibitem [{\citenamefont {Hensen}\ \emph {et~al.}(2015)\citenamefont {Hensen},
  \citenamefont {Bernien}, \citenamefont {Dr{\'e}au}, \citenamefont {Reiserer},
  \citenamefont {Kalb}, \citenamefont {Blok}, \citenamefont {Ruitenberg},
  \citenamefont {Vermeulen}, \citenamefont {Schouten}, \citenamefont
  {Abell{\'a}n}, \citenamefont {Amaya}, \citenamefont {Pruneri}, \citenamefont
  {Mitchell}, \citenamefont {Markham}, \citenamefont {Twitchen}, \citenamefont
  {Elkouss}, \citenamefont {Wehner}, \citenamefont {Taminiau},\ and\
  \citenamefont {Hanson}}]{HensenB:Nature2015}%
  \BibitemOpen
  \bibfield  {author} {\bibinfo {author} {\bibfnamefont {B.}~\bibnamefont
  {Hensen}}, \bibinfo {author} {\bibfnamefont {H.}~\bibnamefont {Bernien}},
  \bibinfo {author} {\bibfnamefont {A.~E.}\ \bibnamefont {Dr{\'e}au}}, \bibinfo
  {author} {\bibfnamefont {A.}~\bibnamefont {Reiserer}}, \bibinfo {author}
  {\bibfnamefont {N.}~\bibnamefont {Kalb}}, \bibinfo {author} {\bibfnamefont
  {M.~S.}\ \bibnamefont {Blok}}, \bibinfo {author} {\bibfnamefont
  {J.}~\bibnamefont {Ruitenberg}}, \bibinfo {author} {\bibfnamefont {R.~F.~L.}\
  \bibnamefont {Vermeulen}}, \bibinfo {author} {\bibfnamefont {R.~N.}\
  \bibnamefont {Schouten}}, \bibinfo {author} {\bibfnamefont {C.}~\bibnamefont
  {Abell{\'a}n}}, \bibinfo {author} {\bibfnamefont {W.}~\bibnamefont {Amaya}},
  \bibinfo {author} {\bibfnamefont {V.}~\bibnamefont {Pruneri}}, \bibinfo
  {author} {\bibfnamefont {M.~W.}\ \bibnamefont {Mitchell}}, \bibinfo {author}
  {\bibfnamefont {M.}~\bibnamefont {Markham}}, \bibinfo {author} {\bibfnamefont
  {D.~J.}\ \bibnamefont {Twitchen}}, \bibinfo {author} {\bibfnamefont
  {D.}~\bibnamefont {Elkouss}}, \bibinfo {author} {\bibfnamefont
  {S.}~\bibnamefont {Wehner}}, \bibinfo {author} {\bibfnamefont {T.~H.}\
  \bibnamefont {Taminiau}}, \ and\ \bibinfo {author} {\bibfnamefont
  {R.}~\bibnamefont {Hanson}},\ }\bibfield  {title} {\enquote {\bibinfo {title}
  {Loophole-free {B}ell inequality violation using electron spins separated by
  $1.3$ kilometres},}\ }\href {http://dx.doi.org/10.1038/nature15759}
  {\bibfield  {journal} {\bibinfo  {journal} {Nature}\ }\textbf {\bibinfo
  {volume} {526}},\ \bibinfo {pages} {682--686} (\bibinfo {year}
  {2015})}\BibitemShut {NoStop}%
\bibitem [{\citenamefont {Casola}\ \emph {et~al.}(2018)\citenamefont {Casola},
  \citenamefont {van~der Sar},\ and\ \citenamefont
  {Yacoby}}]{CasolaF:NatRevMater2018}%
  \BibitemOpen
  \bibfield  {author} {\bibinfo {author} {\bibfnamefont {Francesco}\
  \bibnamefont {Casola}}, \bibinfo {author} {\bibfnamefont {Toeno}\
  \bibnamefont {van~der Sar}}, \ and\ \bibinfo {author} {\bibfnamefont {Amir}\
  \bibnamefont {Yacoby}},\ }\bibfield  {title} {\enquote {\bibinfo {title}
  {Probing condensed matter physics with magnetometry based on nitrogen-vacancy
  centres in diamond},}\ }\href@noop {} {\bibfield  {journal} {\bibinfo
  {journal} {Nat. Rev. Mater.}\ }\textbf {\bibinfo {volume} {3}},\ \bibinfo
  {pages} {17088} (\bibinfo {year} {2018})}\BibitemShut {NoStop}%
\bibitem [{\citenamefont {Thiering}\ and\ \citenamefont
  {Gali}(2018)}]{ThieringG:PhysRevB2018}%
  \BibitemOpen
  \bibfield  {author} {\bibinfo {author} {\bibfnamefont {Gerg\H{o}}\
  \bibnamefont {Thiering}}\ and\ \bibinfo {author} {\bibfnamefont {Adam}\
  \bibnamefont {Gali}},\ }\bibfield  {title} {\enquote {\bibinfo {title}
  {Theory of the optical spin-polarization loop of the nitrogen-vacancy center
  in diamond},}\ }\href {\doibase 10.1103/PhysRevB.98.085207} {\bibfield
  {journal} {\bibinfo  {journal} {Phys. Rev. B}\ }\textbf {\bibinfo {volume}
  {98}},\ \bibinfo {pages} {085207} (\bibinfo {year} {2018})}\BibitemShut
  {NoStop}%
\bibitem [{\citenamefont {Udvarhelyi}\ \emph {et~al.}(2018)\citenamefont
  {Udvarhelyi}, \citenamefont {Shkolnikov}, \citenamefont {Gali}, \citenamefont
  {Burkard},\ and\ \citenamefont {P\'alyi}}]{UdvarhelyiP:PhysRevB2018}%
  \BibitemOpen
  \bibfield  {author} {\bibinfo {author} {\bibfnamefont {P\'eter}\ \bibnamefont
  {Udvarhelyi}}, \bibinfo {author} {\bibfnamefont {V.~O.}\ \bibnamefont
  {Shkolnikov}}, \bibinfo {author} {\bibfnamefont {Adam}\ \bibnamefont {Gali}},
  \bibinfo {author} {\bibfnamefont {Guido}\ \bibnamefont {Burkard}}, \ and\
  \bibinfo {author} {\bibfnamefont {Andr\'as}\ \bibnamefont {P\'alyi}},\
  }\bibfield  {title} {\enquote {\bibinfo {title} {Spin-strain interaction in
  nitrogen-vacancy centers in diamond},}\ }\href {\doibase
  10.1103/PhysRevB.98.075201} {\bibfield  {journal} {\bibinfo  {journal} {Phys.
  Rev. B}\ }\textbf {\bibinfo {volume} {98}},\ \bibinfo {pages} {075201}
  (\bibinfo {year} {2018})}\BibitemShut {NoStop}%
\bibitem [{\citenamefont {Plakhotnik}\ \emph {et~al.}(2015)\citenamefont
  {Plakhotnik}, \citenamefont {Doherty},\ and\ \citenamefont
  {Manson}}]{PlakhotnikT:PhysRevB2015}%
  \BibitemOpen
  \bibfield  {author} {\bibinfo {author} {\bibfnamefont {Taras}\ \bibnamefont
  {Plakhotnik}}, \bibinfo {author} {\bibfnamefont {Marcus~W.}\ \bibnamefont
  {Doherty}}, \ and\ \bibinfo {author} {\bibfnamefont {Neil~B.}\ \bibnamefont
  {Manson}},\ }\bibfield  {title} {\enquote {\bibinfo {title} {Electron-phonon
  processes of the nitrogen-vacancy center in diamond},}\ }\href {\doibase
  10.1103/PhysRevB.92.081203} {\bibfield  {journal} {\bibinfo  {journal} {Phys.
  Rev. B}\ }\textbf {\bibinfo {volume} {92}},\ \bibinfo {pages} {081203}
  (\bibinfo {year} {2015})}\BibitemShut {NoStop}%
\bibitem [{\citenamefont {Iwasaki}\ \emph {et~al.}(2017)\citenamefont
  {Iwasaki}, \citenamefont {Miyamoto}, \citenamefont {Taniguchi}, \citenamefont
  {Siyushev}, \citenamefont {Metsch}, \citenamefont {Jelezko},\ and\
  \citenamefont {Hatano}}]{IwasakiT:PhysRevLett2017}%
  \BibitemOpen
  \bibfield  {author} {\bibinfo {author} {\bibfnamefont {Takayuki}\
  \bibnamefont {Iwasaki}}, \bibinfo {author} {\bibfnamefont {Yoshiyuki}\
  \bibnamefont {Miyamoto}}, \bibinfo {author} {\bibfnamefont {Takashi}\
  \bibnamefont {Taniguchi}}, \bibinfo {author} {\bibfnamefont {Petr}\
  \bibnamefont {Siyushev}}, \bibinfo {author} {\bibfnamefont {Mathias~H.}\
  \bibnamefont {Metsch}}, \bibinfo {author} {\bibfnamefont {Fedor}\
  \bibnamefont {Jelezko}}, \ and\ \bibinfo {author} {\bibfnamefont {Mutsuko}\
  \bibnamefont {Hatano}},\ }\bibfield  {title} {\enquote {\bibinfo {title}
  {Tin-{V}acancy {Q}uantum {E}mitters in {D}iamond},}\ }\href {\doibase
  10.1103/PhysRevLett.119.253601} {\bibfield  {journal} {\bibinfo  {journal}
  {Phys. Rev. Lett.}\ }\textbf {\bibinfo {volume} {119}},\ \bibinfo {pages}
  {253601} (\bibinfo {year} {2017})}\BibitemShut {NoStop}%
\bibitem [{\citenamefont {L\"{u}hmann}\ \emph {et~al.}(2018)\citenamefont
  {L\"{u}hmann}, \citenamefont {Raatz}, \citenamefont {John}, \citenamefont
  {Lesik}, \citenamefont {R\"{o}diger}, \citenamefont {Portail}, \citenamefont
  {Wildanger}, \citenamefont {Klei{\ss}ler}, \citenamefont {Nordlund},
  \citenamefont {Zaitsev}, \citenamefont {Roch}, \citenamefont {Tallaire},
  \citenamefont {Meijer},\ and\ \citenamefont {Pezzagna}}]{LumannT:JPhysD2018}%
  \BibitemOpen
  \bibfield  {author} {\bibinfo {author} {\bibfnamefont {Tobias}\ \bibnamefont
  {L\"{u}hmann}}, \bibinfo {author} {\bibfnamefont {Nicole}\ \bibnamefont
  {Raatz}}, \bibinfo {author} {\bibfnamefont {Roger}\ \bibnamefont {John}},
  \bibinfo {author} {\bibfnamefont {Margarita}\ \bibnamefont {Lesik}}, \bibinfo
  {author} {\bibfnamefont {Jasper}\ \bibnamefont {R\"{o}diger}}, \bibinfo
  {author} {\bibfnamefont {Marc}\ \bibnamefont {Portail}}, \bibinfo {author}
  {\bibfnamefont {Dominik}\ \bibnamefont {Wildanger}}, \bibinfo {author}
  {\bibfnamefont {Felix}\ \bibnamefont {Klei{\ss}ler}}, \bibinfo {author}
  {\bibfnamefont {Kai}\ \bibnamefont {Nordlund}}, \bibinfo {author}
  {\bibfnamefont {Alexander}\ \bibnamefont {Zaitsev}}, \bibinfo {author}
  {\bibfnamefont {Jean-Fran\c{c}ois}\ \bibnamefont {Roch}}, \bibinfo {author}
  {\bibfnamefont {Alexandre}\ \bibnamefont {Tallaire}}, \bibinfo {author}
  {\bibfnamefont {Jan}\ \bibnamefont {Meijer}}, \ and\ \bibinfo {author}
  {\bibfnamefont {S\'{e}bastien}\ \bibnamefont {Pezzagna}},\ }\bibfield
  {title} {\enquote {\bibinfo {title} {Screening and engineering of colour
  centres in diamond},}\ }\href
  {http://stacks.iop.org/0022-3727/51/i=48/a=483002} {\bibfield  {journal}
  {\bibinfo  {journal} {J. Phys. D: Appl. Phys.}\ }\textbf {\bibinfo {volume}
  {51}},\ \bibinfo {pages} {483002} (\bibinfo {year} {2018})}\BibitemShut
  {NoStop}%
\bibitem [{\citenamefont {Londero}\ \emph {et~al.}(2018)\citenamefont
  {Londero}, \citenamefont {Bourgeois}, \citenamefont {Nesladek},\ and\
  \citenamefont {Gali}}]{LonderoE:PhysRevB2018}%
  \BibitemOpen
  \bibfield  {author} {\bibinfo {author} {\bibfnamefont {E.}~\bibnamefont
  {Londero}}, \bibinfo {author} {\bibfnamefont {E.}~\bibnamefont {Bourgeois}},
  \bibinfo {author} {\bibfnamefont {M.}~\bibnamefont {Nesladek}}, \ and\
  \bibinfo {author} {\bibfnamefont {A.}~\bibnamefont {Gali}},\ }\bibfield
  {title} {\enquote {\bibinfo {title} {Identification of nickel-vacancy defects
  by combining experimental and ab initio simulated photocurrent spectra},}\
  }\href {\doibase 10.1103/PhysRevB.97.241202} {\bibfield  {journal} {\bibinfo
  {journal} {Phys. Rev. B}\ }\textbf {\bibinfo {volume} {97}},\ \bibinfo
  {pages} {241202} (\bibinfo {year} {2018})}\BibitemShut {NoStop}%
\bibitem [{\citenamefont {Merson}\ \emph {et~al.}(2013)\citenamefont {Merson},
  \citenamefont {Castelletto}, \citenamefont {Aharonovich}, \citenamefont
  {Turbic}, \citenamefont {Kilpatrick},\ and\ \citenamefont
  {Turnley}}]{MersonTD:OptLett2013}%
  \BibitemOpen
  \bibfield  {author} {\bibinfo {author} {\bibfnamefont {Tobias~D.}\
  \bibnamefont {Merson}}, \bibinfo {author} {\bibfnamefont {Stefania}\
  \bibnamefont {Castelletto}}, \bibinfo {author} {\bibfnamefont {Igor}\
  \bibnamefont {Aharonovich}}, \bibinfo {author} {\bibfnamefont {Alisa}\
  \bibnamefont {Turbic}}, \bibinfo {author} {\bibfnamefont {Trevor~J.}\
  \bibnamefont {Kilpatrick}}, \ and\ \bibinfo {author} {\bibfnamefont {Ann~M.}\
  \bibnamefont {Turnley}},\ }\bibfield  {title} {\enquote {\bibinfo {title}
  {Nanodiamonds with silicon vacancy defects for nontoxic photostable
  fluorescent labeling of neural precursor cells},}\ }\href {\doibase
  10.1364/OL.38.004170} {\bibfield  {journal} {\bibinfo  {journal} {Opt.
  Lett.}\ }\textbf {\bibinfo {volume} {38}},\ \bibinfo {pages} {4170--4173}
  (\bibinfo {year} {2013})}\BibitemShut {NoStop}%
\bibitem [{\citenamefont {Iwasaki}\ \emph {et~al.}(2015)\citenamefont
  {Iwasaki}, \citenamefont {Ishibashi}, \citenamefont {Miyamoto}, \citenamefont
  {Doi}, \citenamefont {Kobayashi}, \citenamefont {Miyazaki}, \citenamefont
  {Tahara}, \citenamefont {Jahnke}, \citenamefont {Rogers}, \citenamefont
  {Naydenov}, \citenamefont {Jelezko}, \citenamefont {Yamasaki}, \citenamefont
  {Nagamachi}, \citenamefont {Inubushi}, \citenamefont {Mizuochi},\ and\
  \citenamefont {Hatano}}]{IwasakiT:SciRep2015}%
  \BibitemOpen
  \bibfield  {author} {\bibinfo {author} {\bibfnamefont {Takayuki}\
  \bibnamefont {Iwasaki}}, \bibinfo {author} {\bibfnamefont {Fumitaka}\
  \bibnamefont {Ishibashi}}, \bibinfo {author} {\bibfnamefont {Yoshiyuki}\
  \bibnamefont {Miyamoto}}, \bibinfo {author} {\bibfnamefont {Yuki}\
  \bibnamefont {Doi}}, \bibinfo {author} {\bibfnamefont {Satoshi}\ \bibnamefont
  {Kobayashi}}, \bibinfo {author} {\bibfnamefont {Takehide}\ \bibnamefont
  {Miyazaki}}, \bibinfo {author} {\bibfnamefont {Kosuke}\ \bibnamefont
  {Tahara}}, \bibinfo {author} {\bibfnamefont {Kay~D.}\ \bibnamefont {Jahnke}},
  \bibinfo {author} {\bibfnamefont {Lachlan~J.}\ \bibnamefont {Rogers}},
  \bibinfo {author} {\bibfnamefont {Boris}\ \bibnamefont {Naydenov}}, \bibinfo
  {author} {\bibfnamefont {Fedor}\ \bibnamefont {Jelezko}}, \bibinfo {author}
  {\bibfnamefont {Satoshi}\ \bibnamefont {Yamasaki}}, \bibinfo {author}
  {\bibfnamefont {Shinji}\ \bibnamefont {Nagamachi}}, \bibinfo {author}
  {\bibfnamefont {Toshiro}\ \bibnamefont {Inubushi}}, \bibinfo {author}
  {\bibfnamefont {Norikazu}\ \bibnamefont {Mizuochi}}, \ and\ \bibinfo {author}
  {\bibfnamefont {Mutsuko}\ \bibnamefont {Hatano}},\ }\bibfield  {title}
  {\enquote {\bibinfo {title} {Germanium-{V}acancy {S}ingle {C}olor {C}enters
  in {D}iamond},}\ }\href {http://dx.doi.org/10.1038/srep12882} {\bibfield
  {journal} {\bibinfo  {journal} {Sci. Rep.}\ }\textbf {\bibinfo {volume}
  {5}},\ \bibinfo {pages} {12882} (\bibinfo {year} {2015})}\BibitemShut
  {NoStop}%
\bibitem [{\citenamefont {Palyanov}\ \emph {et~al.}(2015)\citenamefont
  {Palyanov}, \citenamefont {Kupriyanov}, \citenamefont {Borzdov},\ and\
  \citenamefont {Surovtsev}}]{PalyanovYN:SciRep2015}%
  \BibitemOpen
  \bibfield  {author} {\bibinfo {author} {\bibfnamefont {Yuri~N.}\ \bibnamefont
  {Palyanov}}, \bibinfo {author} {\bibfnamefont {Igor~N.}\ \bibnamefont
  {Kupriyanov}}, \bibinfo {author} {\bibfnamefont {Yuri~M.}\ \bibnamefont
  {Borzdov}}, \ and\ \bibinfo {author} {\bibfnamefont {Nikolay~V.}\
  \bibnamefont {Surovtsev}},\ }\bibfield  {title} {\enquote {\bibinfo {title}
  {Germanium: a new catalyst for diamond synthesis and a new optically active
  impurity in diamond},}\ }\href {\doibase 10.1038/srep14789} {\bibfield
  {journal} {\bibinfo  {journal} {Sci. Rep.}\ }\textbf {\bibinfo {volume}
  {5}},\ \bibinfo {pages} {14789} (\bibinfo {year} {2015})}\BibitemShut
  {NoStop}%
\bibitem [{\citenamefont {Green}\ \emph {et~al.}(2017)\citenamefont {Green},
  \citenamefont {Mottishaw}, \citenamefont {Breeze}, \citenamefont {Edmonds},
  \citenamefont {D'Haenens-Johansson}, \citenamefont {Doherty}, \citenamefont
  {Williams}, \citenamefont {Twitchen},\ and\ \citenamefont
  {Newton}}]{GreenBL:PhysRevLett2017}%
  \BibitemOpen
  \bibfield  {author} {\bibinfo {author} {\bibfnamefont {B.~L.}\ \bibnamefont
  {Green}}, \bibinfo {author} {\bibfnamefont {S.}~\bibnamefont {Mottishaw}},
  \bibinfo {author} {\bibfnamefont {B.~G.}\ \bibnamefont {Breeze}}, \bibinfo
  {author} {\bibfnamefont {A.~M.}\ \bibnamefont {Edmonds}}, \bibinfo {author}
  {\bibfnamefont {U.~F.~S.}\ \bibnamefont {D'Haenens-Johansson}}, \bibinfo
  {author} {\bibfnamefont {M.~W.}\ \bibnamefont {Doherty}}, \bibinfo {author}
  {\bibfnamefont {S.~D.}\ \bibnamefont {Williams}}, \bibinfo {author}
  {\bibfnamefont {D.~J.}\ \bibnamefont {Twitchen}}, \ and\ \bibinfo {author}
  {\bibfnamefont {M.~E.}\ \bibnamefont {Newton}},\ }\bibfield  {title}
  {\enquote {\bibinfo {title} {Neutral {S}ilicon-{V}acancy {C}enter in
  {D}iamond: {S}pin {P}olarization and {L}ifetimes},}\ }\href {\doibase
  10.1103/PhysRevLett.119.096402} {\bibfield  {journal} {\bibinfo  {journal}
  {Phys. Rev. Lett.}\ }\textbf {\bibinfo {volume} {119}},\ \bibinfo {pages}
  {096402} (\bibinfo {year} {2017})}\BibitemShut {NoStop}%
\bibitem [{\citenamefont {H\"{a}u{\ss}ler}\ \emph {et~al.}(2017)\citenamefont
  {H\"{a}u{\ss}ler}, \citenamefont {Thiering}, \citenamefont {Dietrich},
  \citenamefont {Waasem}, \citenamefont {Teraji}, \citenamefont {Isoya},
  \citenamefont {Iwasaki}, \citenamefont {Hatano}, \citenamefont {Jelezko},
  \citenamefont {Gali},\ and\ \citenamefont
  {Kubanek}}]{HausslerS:NewJPhys2017}%
  \BibitemOpen
  \bibfield  {author} {\bibinfo {author} {\bibfnamefont {Stefan}\ \bibnamefont
  {H\"{a}u{\ss}ler}}, \bibinfo {author} {\bibfnamefont {Gerg{\H{o}}}\
  \bibnamefont {Thiering}}, \bibinfo {author} {\bibfnamefont {Andreas}\
  \bibnamefont {Dietrich}}, \bibinfo {author} {\bibfnamefont {Niklas}\
  \bibnamefont {Waasem}}, \bibinfo {author} {\bibfnamefont {Tokuyuki}\
  \bibnamefont {Teraji}}, \bibinfo {author} {\bibfnamefont {Junichi}\
  \bibnamefont {Isoya}}, \bibinfo {author} {\bibfnamefont {Takayuki}\
  \bibnamefont {Iwasaki}}, \bibinfo {author} {\bibfnamefont {Mutsuko}\
  \bibnamefont {Hatano}}, \bibinfo {author} {\bibfnamefont {Fedor}\
  \bibnamefont {Jelezko}}, \bibinfo {author} {\bibfnamefont {Adam}\
  \bibnamefont {Gali}}, \ and\ \bibinfo {author} {\bibfnamefont {Alexander}\
  \bibnamefont {Kubanek}},\ }\bibfield  {title} {\enquote {\bibinfo {title}
  {Photoluminescence excitation spectroscopy of {SiV}$^-$ and {GeV}$^-$ color
  center in diamond},}\ }\href {\doibase 10.1088/1367-2630/aa73e5} {\bibfield
  {journal} {\bibinfo  {journal} {New J. Phys.}\ }\textbf {\bibinfo {volume}
  {19}},\ \bibinfo {pages} {063036} (\bibinfo {year} {2017})}\BibitemShut
  {NoStop}%
\bibitem [{\citenamefont {Jones}(2006)}]{Jones:OptMater2006}%
  \BibitemOpen
  \bibfield  {author} {\bibinfo {author} {\bibfnamefont {R.}~\bibnamefont
  {Jones}},\ }\bibfield  {title} {\enquote {\bibinfo {title} {Structure and
  electrical activity of rare-earth dopants in semiconductors},}\ }\href
  {\doibase https://doi.org/10.1016/j.optmat.2005.09.005} {\bibfield  {journal}
  {\bibinfo  {journal} {Opt. Mater.}\ }\textbf {\bibinfo {volume} {28}},\
  \bibinfo {pages} {718 -- 722} (\bibinfo {year} {2006})}\BibitemShut {NoStop}%
\bibitem [{\citenamefont {Sanna}\ \emph {et~al.}(2009)\citenamefont {Sanna},
  \citenamefont {Schmidt}, \citenamefont {Frauenheim},\ and\ \citenamefont
  {Gerstmann}}]{SanaS:PhysRevB2009}%
  \BibitemOpen
  \bibfield  {author} {\bibinfo {author} {\bibfnamefont {Simone}\ \bibnamefont
  {Sanna}}, \bibinfo {author} {\bibfnamefont {W.~G.}\ \bibnamefont {Schmidt}},
  \bibinfo {author} {\bibfnamefont {Th.}\ \bibnamefont {Frauenheim}}, \ and\
  \bibinfo {author} {\bibfnamefont {U.}~\bibnamefont {Gerstmann}},\ }\bibfield
  {title} {\enquote {\bibinfo {title} {Rare-earth defect pairs in {GaN}:
  {LDA}+{U} calculations},}\ }\href {\doibase 10.1103/PhysRevB.80.104120}
  {\bibfield  {journal} {\bibinfo  {journal} {Phys. Rev. B}\ }\textbf {\bibinfo
  {volume} {80}},\ \bibinfo {pages} {104120} (\bibinfo {year}
  {2009})}\BibitemShut {NoStop}%
\bibitem [{\citenamefont {Ouma}\ and\ \citenamefont
  {Meyer}(2014)}]{OumaCNM:PhysBCondMat2014}%
  \BibitemOpen
  \bibfield  {author} {\bibinfo {author} {\bibfnamefont {Cecil~N.M.}\
  \bibnamefont {Ouma}}\ and\ \bibinfo {author} {\bibfnamefont {Walter~E.}\
  \bibnamefont {Meyer}},\ }\bibfield  {title} {\enquote {\bibinfo {title} {Ab
  initio study of metastability of {Eu}$^{3+}$ defect complexes in {GaN}},}\
  }\href {\doibase https://doi.org/10.1016/j.physb.2013.11.004} {\bibfield
  {journal} {\bibinfo  {journal} {Phys. B: Cond. Matt.}\ }\textbf {\bibinfo
  {volume} {439}},\ \bibinfo {pages} {141 -- 143} (\bibinfo {year}
  {2014})}\BibitemShut {NoStop}%
\bibitem [{\citenamefont {Cajzl}\ \emph {et~al.}(2017)\citenamefont {Cajzl},
  \citenamefont {Nekvindova}, \citenamefont {Mackova}, \citenamefont
  {Malinsky}, \citenamefont {Sedmidubsky}, \citenamefont {Husak}, \citenamefont
  {Remes}, \citenamefont {Varga}, \citenamefont {Kromka}, \citenamefont
  {Bottger},\ and\ \citenamefont {Oswald}}]{CajzlJ:PhysChemChemPhys2017}%
  \BibitemOpen
  \bibfield  {author} {\bibinfo {author} {\bibfnamefont {Jakub}\ \bibnamefont
  {Cajzl}}, \bibinfo {author} {\bibfnamefont {Pavla}\ \bibnamefont
  {Nekvindova}}, \bibinfo {author} {\bibfnamefont {Anna}\ \bibnamefont
  {Mackova}}, \bibinfo {author} {\bibfnamefont {Petr}\ \bibnamefont
  {Malinsky}}, \bibinfo {author} {\bibfnamefont {David}\ \bibnamefont
  {Sedmidubsky}}, \bibinfo {author} {\bibfnamefont {Michal}\ \bibnamefont
  {Husak}}, \bibinfo {author} {\bibfnamefont {Zdenek}\ \bibnamefont {Remes}},
  \bibinfo {author} {\bibfnamefont {Marian}\ \bibnamefont {Varga}}, \bibinfo
  {author} {\bibfnamefont {Alexander}\ \bibnamefont {Kromka}}, \bibinfo
  {author} {\bibfnamefont {Roman}\ \bibnamefont {Bottger}}, \ and\ \bibinfo
  {author} {\bibfnamefont {Jiri}\ \bibnamefont {Oswald}},\ }\bibfield  {title}
  {\enquote {\bibinfo {title} {Erbium ion implantation into diamond --
  measurement and modelling of the crystal structure},}\ }\href {\doibase
  10.1039/C6CP08851A} {\bibfield  {journal} {\bibinfo  {journal} {Phys. Chem.
  Chem. Phys.}\ }\textbf {\bibinfo {volume} {19}},\ \bibinfo {pages}
  {6233--6245} (\bibinfo {year} {2017})}\BibitemShut {NoStop}%
\bibitem [{\citenamefont {Cajzl}\ \emph {et~al.}(2018)\citenamefont {Cajzl},
  \citenamefont {Akhetova}, \citenamefont {Nekvindov\'{a}}, \citenamefont
  {Mackov\'{a}}, \citenamefont {Malinsk\'{y}}, \citenamefont {Oswald},
  \citenamefont {Reme\v{s}}, \citenamefont {Varga},\ and\ \citenamefont
  {Kromka}}]{CajzlJ:SurfInterAnal2018}%
  \BibitemOpen
  \bibfield  {author} {\bibinfo {author} {\bibfnamefont {Jakub}\ \bibnamefont
  {Cajzl}}, \bibinfo {author} {\bibfnamefont {Banu}\ \bibnamefont {Akhetova}},
  \bibinfo {author} {\bibfnamefont {Pavla}\ \bibnamefont {Nekvindov\'{a}}},
  \bibinfo {author} {\bibfnamefont {Anna}\ \bibnamefont {Mackov\'{a}}},
  \bibinfo {author} {\bibfnamefont {Petr}\ \bibnamefont {Malinsk\'{y}}},
  \bibinfo {author} {\bibfnamefont {Ji\v{r}\'{i}}\ \bibnamefont {Oswald}},
  \bibinfo {author} {\bibfnamefont {Zden\v{e}k}\ \bibnamefont {Reme\v{s}}},
  \bibinfo {author} {\bibfnamefont {Mari\'{a}n}\ \bibnamefont {Varga}}, \ and\
  \bibinfo {author} {\bibfnamefont {Alexander}\ \bibnamefont {Kromka}},\
  }\bibfield  {title} {\enquote {\bibinfo {title} {Co-implantation of {Er} and
  {Yb} ions into single-crystalline and nano-crystalline diamond},}\ }\href
  {\doibase 10.1002/sia.6407} {\bibfield  {journal} {\bibinfo  {journal} {Surf.
  Interface Anal.}\ }\textbf {\bibinfo {volume} {50}},\ \bibinfo {pages}
  {1218--1223} (\bibinfo {year} {2018})}\BibitemShut {NoStop}%
\bibitem [{\citenamefont {Tan}\ \emph {et~al.}(2018)\citenamefont {Tan},
  \citenamefont {Liu}, \citenamefont {Liu}, \citenamefont {Ren}, \citenamefont
  {Sun}, \citenamefont {Jia}, \citenamefont {Liu}, \citenamefont {Chen},\ and\
  \citenamefont {Wei}}]{TanX:AIPAdv2018}%
  \BibitemOpen
  \bibfield  {author} {\bibinfo {author} {\bibfnamefont {Xin}\ \bibnamefont
  {Tan}}, \bibinfo {author} {\bibfnamefont {Tiebang}\ \bibnamefont {Liu}},
  \bibinfo {author} {\bibfnamefont {Xuejie}\ \bibnamefont {Liu}}, \bibinfo
  {author} {\bibfnamefont {Yuan}\ \bibnamefont {Ren}}, \bibinfo {author}
  {\bibfnamefont {Shiyang}\ \bibnamefont {Sun}}, \bibinfo {author}
  {\bibfnamefont {Huiling}\ \bibnamefont {Jia}}, \bibinfo {author}
  {\bibfnamefont {Zhixin}\ \bibnamefont {Liu}}, \bibinfo {author}
  {\bibfnamefont {Luhua}\ \bibnamefont {Chen}}, \ and\ \bibinfo {author}
  {\bibfnamefont {Xueyuan}\ \bibnamefont {Wei}},\ }\bibfield  {title} {\enquote
  {\bibinfo {title} {Structural stability of {Pr}-related defects in diamond
  and electronic structure single photon source: {A} first-principles study},}\
  }\href {\doibase 10.1063/1.5050412} {\bibfield  {journal} {\bibinfo
  {journal} {AIP Advances}\ }\textbf {\bibinfo {volume} {8}},\ \bibinfo {pages}
  {105202} (\bibinfo {year} {2018})}\BibitemShut {NoStop}%
\bibitem [{\citenamefont {Chaudhry}\ \emph {et~al.}(2014)\citenamefont
  {Chaudhry}, \citenamefont {Boutchko}, \citenamefont {Chourou}, \citenamefont
  {Zhang}, \citenamefont {Gr\o{}nbech-Jensen},\ and\ \citenamefont
  {Canning}}]{ChaudhryA:PhysRevB2014}%
  \BibitemOpen
  \bibfield  {author} {\bibinfo {author} {\bibfnamefont {A.}~\bibnamefont
  {Chaudhry}}, \bibinfo {author} {\bibfnamefont {R.}~\bibnamefont {Boutchko}},
  \bibinfo {author} {\bibfnamefont {S.}~\bibnamefont {Chourou}}, \bibinfo
  {author} {\bibfnamefont {G.}~\bibnamefont {Zhang}}, \bibinfo {author}
  {\bibfnamefont {N.}~\bibnamefont {Gr\o{}nbech-Jensen}}, \ and\ \bibinfo
  {author} {\bibfnamefont {A.}~\bibnamefont {Canning}},\ }\bibfield  {title}
  {\enquote {\bibinfo {title} {First-principles study of luminescence in
  {Eu}${}^{2+}$-doped inorganic scintillators},}\ }\href {\doibase
  10.1103/PhysRevB.89.155105} {\bibfield  {journal} {\bibinfo  {journal} {Phys.
  Rev. B}\ }\textbf {\bibinfo {volume} {89}},\ \bibinfo {pages} {155105}
  (\bibinfo {year} {2014})}\BibitemShut {NoStop}%
\bibitem [{\citenamefont {Dhara}\ \emph {et~al.}(2014)\citenamefont {Dhara},
  \citenamefont {Imakita}, \citenamefont {Mizuhata},\ and\ \citenamefont
  {Fujii}}]{DharaS:Nanotech2014}%
  \BibitemOpen
  \bibfield  {author} {\bibinfo {author} {\bibfnamefont {Soumen}\ \bibnamefont
  {Dhara}}, \bibinfo {author} {\bibfnamefont {Kenji}\ \bibnamefont {Imakita}},
  \bibinfo {author} {\bibfnamefont {Minoru}\ \bibnamefont {Mizuhata}}, \ and\
  \bibinfo {author} {\bibfnamefont {Minoru}\ \bibnamefont {Fujii}},\ }\bibfield
   {title} {\enquote {\bibinfo {title} {Europium doping induced symmetry
  deviation and its impact on the second harmonic generation of doped {ZnO}
  nanowires},}\ }\href {http://stacks.iop.org/0957-4484/25/i=22/a=225202}
  {\bibfield  {journal} {\bibinfo  {journal} {Nanotechnology}\ }\textbf
  {\bibinfo {volume} {25}},\ \bibinfo {pages} {225202} (\bibinfo {year}
  {2014})}\BibitemShut {NoStop}%
\bibitem [{\citenamefont
  {Vanpoucke}(2014)}]{VanpouckeDannyEP:2014f_Nanotechnology}%
  \BibitemOpen
  \bibfield  {author} {\bibinfo {author} {\bibfnamefont {Danny E.~P.}\
  \bibnamefont {Vanpoucke}},\ }\bibfield  {title} {\enquote {\bibinfo {title}
  {Comment on '{E}uropium doping induced symmetry deviation and its impact on
  the second harmonic generation of doped {ZnO} nanowires'},}\ }\href {\doibase
  10.1088/0957-4484/25/45/458001} {\bibfield  {journal} {\bibinfo  {journal}
  {Nanotechnology}\ }\textbf {\bibinfo {volume} {25}},\ \bibinfo {pages}
  {458001} (\bibinfo {year} {2014})}\BibitemShut {NoStop}%
\bibitem [{\citenamefont {Binnemans}(2015)}]{BinnemansK:CoordChemRev2015}%
  \BibitemOpen
  \bibfield  {author} {\bibinfo {author} {\bibfnamefont {Koen}\ \bibnamefont
  {Binnemans}},\ }\bibfield  {title} {\enquote {\bibinfo {title}
  {Interpretation of europium(iii) spectra},}\ }\href {\doibase
  10.1016/j.ccr.2015.02.015} {\bibfield  {journal} {\bibinfo  {journal} {Coord.
  Chem. Rev.}\ }\textbf {\bibinfo {volume} {295}},\ \bibinfo {pages} {1 -- 45}
  (\bibinfo {year} {2015})}\BibitemShut {NoStop}%
\bibitem [{\citenamefont {Gupta}\ \emph {et~al.}(2016)\citenamefont {Gupta},
  \citenamefont {Sudarshan}, \citenamefont {Ghosh}, \citenamefont {Sanyal},
  \citenamefont {Srivastava}, \citenamefont {Arya}, \citenamefont {Pujari},\
  and\ \citenamefont {Kadam}}]{GuptaSK:RSCAdv2016}%
  \BibitemOpen
  \bibfield  {author} {\bibinfo {author} {\bibfnamefont {Santosh~K.}\
  \bibnamefont {Gupta}}, \bibinfo {author} {\bibfnamefont {K.}~\bibnamefont
  {Sudarshan}}, \bibinfo {author} {\bibfnamefont {P.~S.}\ \bibnamefont
  {Ghosh}}, \bibinfo {author} {\bibfnamefont {K.}~\bibnamefont {Sanyal}},
  \bibinfo {author} {\bibfnamefont {A.~P.}\ \bibnamefont {Srivastava}},
  \bibinfo {author} {\bibfnamefont {A.}~\bibnamefont {Arya}}, \bibinfo {author}
  {\bibfnamefont {P.~K.}\ \bibnamefont {Pujari}}, \ and\ \bibinfo {author}
  {\bibfnamefont {R.~M.}\ \bibnamefont {Kadam}},\ }\bibfield  {title} {\enquote
  {\bibinfo {title} {Luminescence of undoped and {Eu}$^{3+}$ doped
  nanocrystalline {SrWO}$_4$ scheelite: time resolved fluorescence complimented
  by {DFT} and positron annihilation spectroscopic studies},}\ }\href {\doibase
  10.1039/C5RA23876E} {\bibfield  {journal} {\bibinfo  {journal} {RSC Adv.}\
  }\textbf {\bibinfo {volume} {6}},\ \bibinfo {pages} {3792--3805} (\bibinfo
  {year} {2016})}\BibitemShut {NoStop}%
\bibitem [{\citenamefont {Lorke}\ \emph {et~al.}(2016)\citenamefont {Lorke},
  \citenamefont {Frauenheim},\ and\ \citenamefont
  {da~Rosa}}]{LorkeM:PhysRevB2016}%
  \BibitemOpen
  \bibfield  {author} {\bibinfo {author} {\bibfnamefont {M.}~\bibnamefont
  {Lorke}}, \bibinfo {author} {\bibfnamefont {T.}~\bibnamefont {Frauenheim}}, \
  and\ \bibinfo {author} {\bibfnamefont {A.~L.}\ \bibnamefont {da~Rosa}},\
  }\bibfield  {title} {\enquote {\bibinfo {title} {Many-body electronic
  structure calculations of {Eu}-doped {ZnO}},}\ }\href {\doibase
  10.1103/PhysRevB.93.115132} {\bibfield  {journal} {\bibinfo  {journal} {Phys.
  Rev. B}\ }\textbf {\bibinfo {volume} {93}},\ \bibinfo {pages} {115132}
  (\bibinfo {year} {2016})}\BibitemShut {NoStop}%
\bibitem [{\citenamefont {Yang}\ \emph {et~al.}(2016)\citenamefont {Yang},
  \citenamefont {Wang}, \citenamefont {Cormack}, \citenamefont {Zych},
  \citenamefont {Seo},\ and\ \citenamefont {Wu}}]{YangY:OptMaterExpress2016}%
  \BibitemOpen
  \bibfield  {author} {\bibinfo {author} {\bibfnamefont {Yan}\ \bibnamefont
  {Yang}}, \bibinfo {author} {\bibfnamefont {Bu}~\bibnamefont {Wang}}, \bibinfo
  {author} {\bibfnamefont {Alastair}\ \bibnamefont {Cormack}}, \bibinfo
  {author} {\bibfnamefont {Eugeniusz}\ \bibnamefont {Zych}}, \bibinfo {author}
  {\bibfnamefont {Hyo~Jin}\ \bibnamefont {Seo}}, \ and\ \bibinfo {author}
  {\bibfnamefont {Yiquan}\ \bibnamefont {Wu}},\ }\bibfield  {title} {\enquote
  {\bibinfo {title} {Theoretical analysis and experiment on eu reduction in
  alumina optical materials},}\ }\href {\doibase 10.1364/OME.6.002404}
  {\bibfield  {journal} {\bibinfo  {journal} {Opt. Mater. Express}\ }\textbf
  {\bibinfo {volume} {6}},\ \bibinfo {pages} {2404--2412} (\bibinfo {year}
  {2016})}\BibitemShut {NoStop}%
\bibitem [{\citenamefont {Ekimov}\ \emph
  {et~al.}(2017{\natexlab{a}})\citenamefont {Ekimov}, \citenamefont {Zibrov},
  \citenamefont {Malykhin}, \citenamefont {Khmelnitskiy},\ and\ \citenamefont
  {Vlasov}}]{Ekimov:MatterLett2017}%
  \BibitemOpen
  \bibfield  {author} {\bibinfo {author} {\bibfnamefont {E.A.}\ \bibnamefont
  {Ekimov}}, \bibinfo {author} {\bibfnamefont {I.P.}\ \bibnamefont {Zibrov}},
  \bibinfo {author} {\bibfnamefont {S.A.}\ \bibnamefont {Malykhin}}, \bibinfo
  {author} {\bibfnamefont {R.A.}\ \bibnamefont {Khmelnitskiy}}, \ and\ \bibinfo
  {author} {\bibfnamefont {I.I.}\ \bibnamefont {Vlasov}},\ }\bibfield  {title}
  {\enquote {\bibinfo {title} {Synthesis of diamond in double carbon-rare earth
  element systems},}\ }\href {\doibase
  https://doi.org/10.1016/j.matlet.2017.01.110} {\bibfield  {journal} {\bibinfo
   {journal} {Mater. Lett.}\ }\textbf {\bibinfo {volume} {193}},\ \bibinfo
  {pages} {130 -- 132} (\bibinfo {year} {2017}{\natexlab{a}})}\BibitemShut
  {NoStop}%
\bibitem [{\citenamefont {Ekimov}\ \emph
  {et~al.}(2017{\natexlab{b}})\citenamefont {Ekimov}, \citenamefont {Zibrov},
  \citenamefont {Malykhin}, \citenamefont {Khmel'nitskii},\ and\ \citenamefont
  {Vlasov}}]{EkimovEA:InorgMater2017}%
  \BibitemOpen
  \bibfield  {author} {\bibinfo {author} {\bibfnamefont {E.~A.}\ \bibnamefont
  {Ekimov}}, \bibinfo {author} {\bibfnamefont {I.~P.}\ \bibnamefont {Zibrov}},
  \bibinfo {author} {\bibfnamefont {S.~A.}\ \bibnamefont {Malykhin}}, \bibinfo
  {author} {\bibfnamefont {R.~A.}\ \bibnamefont {Khmel'nitskii}}, \ and\
  \bibinfo {author} {\bibfnamefont {I.~I.}\ \bibnamefont {Vlasov}},\ }\bibfield
   {title} {\enquote {\bibinfo {title} {Luminescence properties of diamond
  prepared in the presence of rare-earth elements},}\ }\href {\doibase
  10.1134/S0020168517080039} {\bibfield  {journal} {\bibinfo  {journal} {Inorg.
  Mater.}\ }\textbf {\bibinfo {volume} {53}},\ \bibinfo {pages} {809--815}
  (\bibinfo {year} {2017}{\natexlab{b}})}\BibitemShut {NoStop}%
\bibitem [{\citenamefont {Magyar}\ \emph {et~al.}(2014)\citenamefont {Magyar},
  \citenamefont {Hu}, \citenamefont {Shanley}, \citenamefont {Flatt\'{e}},
  \citenamefont {Hu},\ and\ \citenamefont {Aharonovich}}]{MagyarA:NatComm2014}%
  \BibitemOpen
  \bibfield  {author} {\bibinfo {author} {\bibfnamefont {Andrew}\ \bibnamefont
  {Magyar}}, \bibinfo {author} {\bibfnamefont {Wenhao}\ \bibnamefont {Hu}},
  \bibinfo {author} {\bibfnamefont {Toby}\ \bibnamefont {Shanley}}, \bibinfo
  {author} {\bibfnamefont {Michael~E.}\ \bibnamefont {Flatt\'{e}}}, \bibinfo
  {author} {\bibfnamefont {Evelyn}\ \bibnamefont {Hu}}, \ and\ \bibinfo
  {author} {\bibfnamefont {Igor}\ \bibnamefont {Aharonovich}},\ }\bibfield
  {title} {\enquote {\bibinfo {title} {Synthesis of luminescent europium
  defects in diamond},}\ }\href {\doibase 10.1038/ncomms4523} {\bibfield
  {journal} {\bibinfo  {journal} {Nat. Commun.}\ }\textbf {\bibinfo {volume}
  {5}},\ \bibinfo {pages} {3523} (\bibinfo {year} {2014})}\BibitemShut
  {NoStop}%
\bibitem [{\citenamefont {Sedov}\ \emph {et~al.}(2017)\citenamefont {Sedov},
  \citenamefont {Kuznetsov}, \citenamefont {Ralchenko}, \citenamefont
  {Mayakova}, \citenamefont {Krivobok}, \citenamefont {Savin}, \citenamefont
  {Zhuravlev}, \citenamefont {Martyanov}, \citenamefont {Romanishkin},
  \citenamefont {Khomich}, \citenamefont {Fedorov},\ and\ \citenamefont
  {Konov}}]{SedovV:DRM2017}%
  \BibitemOpen
  \bibfield  {author} {\bibinfo {author} {\bibfnamefont {V.S.}\ \bibnamefont
  {Sedov}}, \bibinfo {author} {\bibfnamefont {S.V.}\ \bibnamefont {Kuznetsov}},
  \bibinfo {author} {\bibfnamefont {V.G.}\ \bibnamefont {Ralchenko}}, \bibinfo
  {author} {\bibfnamefont {M.N.}\ \bibnamefont {Mayakova}}, \bibinfo {author}
  {\bibfnamefont {V.S.}\ \bibnamefont {Krivobok}}, \bibinfo {author}
  {\bibfnamefont {S.S.}\ \bibnamefont {Savin}}, \bibinfo {author}
  {\bibfnamefont {K.P.}\ \bibnamefont {Zhuravlev}}, \bibinfo {author}
  {\bibfnamefont {A.K.}\ \bibnamefont {Martyanov}}, \bibinfo {author}
  {\bibfnamefont {I.D.}\ \bibnamefont {Romanishkin}}, \bibinfo {author}
  {\bibfnamefont {A.A.}\ \bibnamefont {Khomich}}, \bibinfo {author}
  {\bibfnamefont {P.P.}\ \bibnamefont {Fedorov}}, \ and\ \bibinfo {author}
  {\bibfnamefont {V.I.}\ \bibnamefont {Konov}},\ }\bibfield  {title} {\enquote
  {\bibinfo {title} {Diamond-{EuF}$_3$ nanocomposites with bright orange
  photoluminescence},}\ }\href {\doibase
  https://doi.org/10.1016/j.diamond.2016.12.022} {\bibfield  {journal}
  {\bibinfo  {journal} {Diam. Relat. Mater.}\ }\textbf {\bibinfo {volume}
  {72}},\ \bibinfo {pages} {47 -- 52} (\bibinfo {year} {2017})}\BibitemShut
  {NoStop}%
\bibitem [{\citenamefont {Momma}\ and\ \citenamefont
  {Izumi}(2008)}]{VESTA:JApplCryst2008}%
  \BibitemOpen
  \bibfield  {author} {\bibinfo {author} {\bibfnamefont {Koichi}\ \bibnamefont
  {Momma}}\ and\ \bibinfo {author} {\bibfnamefont {Fujio}\ \bibnamefont
  {Izumi}},\ }\bibfield  {title} {\enquote {\bibinfo {title} {{VESTA}: a
  three-dimensional visualization system for electronic and structural
  analysis},}\ }\href {\doibase 10.1107/S0021889808012016} {\bibfield
  {journal} {\bibinfo  {journal} {J. Appl. Cryst.}\ }\textbf {\bibinfo {volume}
  {41}},\ \bibinfo {pages} {653--658} (\bibinfo {year} {2008})}\BibitemShut
  {NoStop}%
\bibitem [{\citenamefont {Perdew}\ \emph {et~al.}(1996)\citenamefont {Perdew},
  \citenamefont {Burke},\ and\ \citenamefont {Ernzerhof}}]{PBE_1996prl}%
  \BibitemOpen
  \bibfield  {author} {\bibinfo {author} {\bibfnamefont {John~P.}\ \bibnamefont
  {Perdew}}, \bibinfo {author} {\bibfnamefont {Kieron}\ \bibnamefont {Burke}},
  \ and\ \bibinfo {author} {\bibfnamefont {Matthias}\ \bibnamefont
  {Ernzerhof}},\ }\bibfield  {title} {\enquote {\bibinfo {title} {Generalized
  gradient approximation made simple},}\ }\href {\doibase
  10.1103/PhysRevLett.77.3865} {\bibfield  {journal} {\bibinfo  {journal}
  {Phys. Rev. Lett.}\ }\textbf {\bibinfo {volume} {77}},\ \bibinfo {pages}
  {3865--3868} (\bibinfo {year} {1996})}\BibitemShut {NoStop}%
\bibitem [{\citenamefont {Liechtenstein}\ \emph {et~al.}(1995)\citenamefont
  {Liechtenstein}, \citenamefont {Anisimov},\ and\ \citenamefont
  {Zaanen}}]{LiechtensteinAI:PhysRevB1995}%
  \BibitemOpen
  \bibfield  {author} {\bibinfo {author} {\bibfnamefont {A.~I.}\ \bibnamefont
  {Liechtenstein}}, \bibinfo {author} {\bibfnamefont {V.~I.}\ \bibnamefont
  {Anisimov}}, \ and\ \bibinfo {author} {\bibfnamefont {J.}~\bibnamefont
  {Zaanen}},\ }\bibfield  {title} {\enquote {\bibinfo {title}
  {Density-functional theory and strong interactions: {O}rbital ordering in
  {M}ott-{H}ubbard insulators},}\ }\href {\doibase 10.1103/PhysRevB.52.R5467}
  {\bibfield  {journal} {\bibinfo  {journal} {Phys. Rev. B}\ }\textbf {\bibinfo
  {volume} {52}},\ \bibinfo {pages} {R5467--R5470} (\bibinfo {year}
  {1995})}\BibitemShut {NoStop}%
\bibitem [{\citenamefont {Johannes}\ and\ \citenamefont
  {Pickett}(2005)}]{JohannesMD:PhysRevB2005}%
  \BibitemOpen
  \bibfield  {author} {\bibinfo {author} {\bibfnamefont {M.~D.}\ \bibnamefont
  {Johannes}}\ and\ \bibinfo {author} {\bibfnamefont {W.~E.}\ \bibnamefont
  {Pickett}},\ }\bibfield  {title} {\enquote {\bibinfo {title} {Magnetic
  coupling between nonmagnetic ions: {Eu}$^{3+}$ in {EuN} and {EuP}},}\ }\href
  {\doibase 10.1103/PhysRevB.72.195116} {\bibfield  {journal} {\bibinfo
  {journal} {Phys. Rev. B}\ }\textbf {\bibinfo {volume} {72}},\ \bibinfo
  {pages} {195116} (\bibinfo {year} {2005})}\BibitemShut {NoStop}%
\bibitem [{\citenamefont {Larson}\ and\ \citenamefont
  {Lambrecht}(2006)}]{LarsonP:JPhysCondMat2006}%
  \BibitemOpen
  \bibfield  {author} {\bibinfo {author} {\bibfnamefont {Paul}\ \bibnamefont
  {Larson}}\ and\ \bibinfo {author} {\bibfnamefont {Walter R~L}\ \bibnamefont
  {Lambrecht}},\ }\bibfield  {title} {\enquote {\bibinfo {title} {Electronic
  structure and magnetism of europium chalcogenides in comparison with
  gadolinium nitride},}\ }\href {http://stacks.iop.org/0953-8984/18/i=49/a=024}
  {\bibfield  {journal} {\bibinfo  {journal} {J. Phys.: Cond. Matter}\ }\textbf
  {\bibinfo {volume} {18}},\ \bibinfo {pages} {11333} (\bibinfo {year}
  {2006})}\BibitemShut {NoStop}%
\bibitem [{\citenamefont {Vanpoucke}\ and\ \citenamefont
  {Haenen}(2017)}]{VanpouckeDannyEP:2017d_DRM_DFTUVac}%
  \BibitemOpen
  \bibfield  {author} {\bibinfo {author} {\bibfnamefont {Danny E.~P.}\
  \bibnamefont {Vanpoucke}}\ and\ \bibinfo {author} {\bibfnamefont {Ken}\
  \bibnamefont {Haenen}},\ }\bibfield  {title} {\enquote {\bibinfo {title}
  {Revisiting the neutral {C}-vacancy in diamond: {L}ocalization of electrons
  through {DFT+U}},}\ }\href {\doibase 10.1016/j.diamond.2017.08.009}
  {\bibfield  {journal} {\bibinfo  {journal} {Diam. Relat. Mater.}\ }\textbf
  {\bibinfo {volume} {79}},\ \bibinfo {pages} {60--69} (\bibinfo {year}
  {2017})}\BibitemShut {NoStop}%
\bibitem [{\citenamefont {Heyd}\ \emph {et~al.}(2003)\citenamefont {Heyd},
  \citenamefont {Scuseria},\ and\ \citenamefont
  {Ernzerhof}}]{HeydJ:JChemPhys2003_HSE}%
  \BibitemOpen
  \bibfield  {author} {\bibinfo {author} {\bibfnamefont {Jochen}\ \bibnamefont
  {Heyd}}, \bibinfo {author} {\bibfnamefont {Gustavo~E.}\ \bibnamefont
  {Scuseria}}, \ and\ \bibinfo {author} {\bibfnamefont {Matthias}\ \bibnamefont
  {Ernzerhof}},\ }\bibfield  {title} {\enquote {\bibinfo {title} {Hybrid
  functionals based on a screened {C}oulomb potential},}\ }\href {\doibase
  http://dx.doi.org/10.1063/1.1564060} {\bibfield  {journal} {\bibinfo
  {journal} {J. Chem. Phys.}\ }\textbf {\bibinfo {volume} {118}},\ \bibinfo
  {pages} {8207--8215} (\bibinfo {year} {2003})}\BibitemShut {NoStop}%
\bibitem [{\citenamefont {Heyd}\ \emph {et~al.}(2005)\citenamefont {Heyd},
  \citenamefont {Peralta}, \citenamefont {Scuseria},\ and\ \citenamefont
  {Martin}}]{Heyd:JChemPhys2005_HSE_BG}%
  \BibitemOpen
  \bibfield  {author} {\bibinfo {author} {\bibfnamefont {Jochen}\ \bibnamefont
  {Heyd}}, \bibinfo {author} {\bibfnamefont {Juan~E.}\ \bibnamefont {Peralta}},
  \bibinfo {author} {\bibfnamefont {Gustavo~E.}\ \bibnamefont {Scuseria}}, \
  and\ \bibinfo {author} {\bibfnamefont {Richard~L.}\ \bibnamefont {Martin}},\
  }\bibfield  {title} {\enquote {\bibinfo {title} {Energy band gaps and lattice
  parameters evaluated with the {Heyd-Scuseria-Ernzerhof} screened hybrid
  functional},}\ }\href {\doibase http://dx.doi.org/10.1063/1.2085170}
  {\bibfield  {journal} {\bibinfo  {journal} {J. Chem. Phys.}\ }\textbf
  {\bibinfo {volume} {123}},\ \bibinfo {eid} {174101} (\bibinfo {year}
  {2005})}\BibitemShut {NoStop}%
\bibitem [{\citenamefont {Henderson}\ \emph {et~al.}(2011)\citenamefont
  {Henderson}, \citenamefont {Paier},\ and\ \citenamefont
  {Scuseria}}]{HendersonTM:PhysStatusSolidiB2011}%
  \BibitemOpen
  \bibfield  {author} {\bibinfo {author} {\bibfnamefont {Thomas~M.}\
  \bibnamefont {Henderson}}, \bibinfo {author} {\bibfnamefont {Joachim}\
  \bibnamefont {Paier}}, \ and\ \bibinfo {author} {\bibfnamefont {Gustavo~E.}\
  \bibnamefont {Scuseria}},\ }\bibfield  {title} {\enquote {\bibinfo {title}
  {Accurate treatment of solids with the {HSE} screened hybrid},}\ }\href
  {\doibase 10.1002/pssb.201046303} {\bibfield  {journal} {\bibinfo  {journal}
  {Phys. status solidi B}\ }\textbf {\bibinfo {volume} {248}},\ \bibinfo
  {pages} {767--774} (\bibinfo {year} {2011})}\BibitemShut {NoStop}%
\bibitem [{\citenamefont {Garza}\ and\ \citenamefont
  {Scuseria}(2016)}]{GarzaA:JPhysChemLett2016}%
  \BibitemOpen
  \bibfield  {author} {\bibinfo {author} {\bibfnamefont {Alejandro~J.}\
  \bibnamefont {Garza}}\ and\ \bibinfo {author} {\bibfnamefont {Gustavo~E.}\
  \bibnamefont {Scuseria}},\ }\bibfield  {title} {\enquote {\bibinfo {title}
  {{P}redicting {B}and {G}aps with {H}ybrid {D}ensity {F}unctionals},}\ }\href
  {\doibase 10.1021/acs.jpclett.6b01807} {\bibfield  {journal} {\bibinfo
  {journal} {J. Phys. Chem. Lett.}\ }\textbf {\bibinfo {volume} {7}},\ \bibinfo
  {pages} {4165--4170} (\bibinfo {year} {2016})}\BibitemShut {NoStop}%
\bibitem [{\citenamefont {Hendrickx}\ \emph {et~al.}(2015)\citenamefont
  {Hendrickx}, \citenamefont {Vanpoucke}, \citenamefont {Leus}, \citenamefont
  {Lejaeghere}, \citenamefont {Deyne}, \citenamefont {Speybroeck},
  \citenamefont {Voort},\ and\ \citenamefont
  {Hemelsoet}}]{HendrickxVanpoucke:InorgChem2015}%
  \BibitemOpen
  \bibfield  {author} {\bibinfo {author} {\bibfnamefont {Kevin}\ \bibnamefont
  {Hendrickx}}, \bibinfo {author} {\bibfnamefont {Danny E.~P.}\ \bibnamefont
  {Vanpoucke}}, \bibinfo {author} {\bibfnamefont {Karen}\ \bibnamefont {Leus}},
  \bibinfo {author} {\bibfnamefont {Kurt}\ \bibnamefont {Lejaeghere}}, \bibinfo
  {author} {\bibfnamefont {Andy Van Yperen-De}\ \bibnamefont {Deyne}}, \bibinfo
  {author} {\bibfnamefont {Veronique~Van}\ \bibnamefont {Speybroeck}}, \bibinfo
  {author} {\bibfnamefont {Pascal Van~Der}\ \bibnamefont {Voort}}, \ and\
  \bibinfo {author} {\bibfnamefont {Karen}\ \bibnamefont {Hemelsoet}},\
  }\bibfield  {title} {\enquote {\bibinfo {title} {{U}nderstanding {I}ntrinsic
  {L}ight {A}bsorption {P}roperties of {UiO-66 F}rameworks: {A C}ombined
  {T}heoretical and {E}xperimental {S}tudy},}\ }\href {\doibase
  10.1021/acs.inorgchem.5b01593} {\bibfield  {journal} {\bibinfo  {journal}
  {Inorg. Chem.}\ }\textbf {\bibinfo {volume} {54}},\ \bibinfo {pages}
  {10701--10710} (\bibinfo {year} {2015})}\BibitemShut {NoStop}%
\bibitem [{\citenamefont
  {Vanpoucke}(2017)}]{VanpouckeDannyEP:2017a_JPhysChemC}%
  \BibitemOpen
  \bibfield  {author} {\bibinfo {author} {\bibfnamefont {Danny E.~P.}\
  \bibnamefont {Vanpoucke}},\ }\bibfield  {title} {\enquote {\bibinfo {title}
  {{L}inker {F}unctionalization in {MIL}-47({V})-{R M}etal–{O}rganic
  {F}rameworks: {U}nderstanding the {E}lectronic {S}tructure},}\ }\href
  {\doibase 10.1021/acs.jpcc.7b01491} {\bibfield  {journal} {\bibinfo
  {journal} {J. Phys. Chem. C}\ }\textbf {\bibinfo {volume} {121}},\ \bibinfo
  {pages} {8014--8022} (\bibinfo {year} {2017})}\BibitemShut {NoStop}%
\bibitem [{\citenamefont {Czelej}\ \emph {et~al.}(2017)\citenamefont {Czelej},
  \citenamefont {\'{S}piewak},\ and\ \citenamefont
  {Kurzyd{\l}owski}}]{CzelejK:DRM2017}%
  \BibitemOpen
  \bibfield  {author} {\bibinfo {author} {\bibfnamefont {K.}~\bibnamefont
  {Czelej}}, \bibinfo {author} {\bibfnamefont {P.}~\bibnamefont {\'{S}piewak}},
  \ and\ \bibinfo {author} {\bibfnamefont {K.~J.}\ \bibnamefont
  {Kurzyd{\l}owski}},\ }\bibfield  {title} {\enquote {\bibinfo {title}
  {Electronic structure of substitutionally doped diamond: {S}pin-polarized,
  hybrid density functional theory analysis},}\ }\href {\doibase
  10.1016/j.diamond.2017.03.009} {\bibfield  {journal} {\bibinfo  {journal}
  {Diam. Rel. Mater.}\ }\textbf {\bibinfo {volume} {75}},\ \bibinfo {pages}
  {146--151} (\bibinfo {year} {2017})}\BibitemShut {NoStop}%
\bibitem [{fn:({\natexlab{a}})}]{fn:forces}%
  \BibitemOpen
  \href@noop {} {} ({\natexlab{a}}),\ \bibinfo {note} {in most systems, the
  largest forces are even below $0.5$meV/\AA.}\BibitemShut {Stop}%
\bibitem [{\citenamefont {Vanpoucke}\ \emph
  {et~al.}(2013{\natexlab{a}})\citenamefont {Vanpoucke}, \citenamefont
  {Bultinck},\ and\ \citenamefont
  {Van~Driessche}}]{VanpouckeDannyEP:2013aJComputChem}%
  \BibitemOpen
  \bibfield  {author} {\bibinfo {author} {\bibfnamefont {Danny E.~P.}\
  \bibnamefont {Vanpoucke}}, \bibinfo {author} {\bibfnamefont {Patrick}\
  \bibnamefont {Bultinck}}, \ and\ \bibinfo {author} {\bibfnamefont {Isabel}\
  \bibnamefont {Van~Driessche}},\ }\bibfield  {title} {\enquote {\bibinfo
  {title} {{Extending Hirshfeld-I to bulk and periodic materials}},}\ }\href
  {\doibase 10.1002/jcc.23088} {\bibfield  {journal} {\bibinfo  {journal} {J.
  Comput. Chem.}\ }\textbf {\bibinfo {volume} {34}},\ \bibinfo {pages}
  {405--417} (\bibinfo {year} {2013}{\natexlab{a}})}\BibitemShut {NoStop}%
\bibitem [{\citenamefont {Vanpoucke}\ \emph
  {et~al.}(2013{\natexlab{b}})\citenamefont {Vanpoucke}, \citenamefont
  {Van~Driessche},\ and\ \citenamefont
  {Bultinck}}]{VanpouckeDannyEP:2013bJComputChem}%
  \BibitemOpen
  \bibfield  {author} {\bibinfo {author} {\bibfnamefont {Danny E.~P.}\
  \bibnamefont {Vanpoucke}}, \bibinfo {author} {\bibfnamefont {Isabel}\
  \bibnamefont {Van~Driessche}}, \ and\ \bibinfo {author} {\bibfnamefont
  {Patrick}\ \bibnamefont {Bultinck}},\ }\bibfield  {title} {\enquote {\bibinfo
  {title} {Reply to `{C}omment on ``{E}xtending {H}irshfeld-{I} to bulk and
  periodic materials'' '},}\ }\href {\doibase 10.1002/jcc.23193} {\bibfield
  {journal} {\bibinfo  {journal} {J. Comput. Chem.}\ }\textbf {\bibinfo
  {volume} {34}},\ \bibinfo {pages} {422--427} (\bibinfo {year}
  {2013}{\natexlab{b}})}\BibitemShut {NoStop}%
\bibitem [{\citenamefont {Wolffis}\ \emph {et~al.}(2019)\citenamefont
  {Wolffis}, \citenamefont {Vanpoucke}, \citenamefont {Sharma}, \citenamefont
  {Lawler},\ and\ \citenamefont
  {Forster}}]{VanpouckeDannyEP:2019b_MicroMesoZeo}%
  \BibitemOpen
  \bibfield  {author} {\bibinfo {author} {\bibfnamefont {Jarod~J.}\
  \bibnamefont {Wolffis}}, \bibinfo {author} {\bibfnamefont {Danny E.~P.}\
  \bibnamefont {Vanpoucke}}, \bibinfo {author} {\bibfnamefont {Amit}\
  \bibnamefont {Sharma}}, \bibinfo {author} {\bibfnamefont {Keith~V.}\
  \bibnamefont {Lawler}}, \ and\ \bibinfo {author} {\bibfnamefont {Paul~M.}\
  \bibnamefont {Forster}},\ }\bibfield  {title} {\enquote {\bibinfo {title}
  {Predicting {P}artial {A}tomic {C}harges in {S}iliceous {Z}eolites},}\ }\href
  {\doibase 10.1016/j.micromeso.2018.10.028} {\bibfield  {journal} {\bibinfo
  {journal} {Microporous Mesoporous Mater.}\ }\textbf {\bibinfo {volume}
  {277}},\ \bibinfo {pages} {184--196} (\bibinfo {year} {2019})}\BibitemShut
  {NoStop}%
\bibitem [{\citenamefont {Lebedev}\ and\ \citenamefont
  {Laikov}(1999)}]{LebedevVI_grid:1999DokladyMath}%
  \BibitemOpen
  \bibfield  {author} {\bibinfo {author} {\bibfnamefont {Vyacheslav~Ivanovich}\
  \bibnamefont {Lebedev}}\ and\ \bibinfo {author} {\bibfnamefont
  {DN}~\bibnamefont {Laikov}},\ }\bibfield  {title} {\enquote {\bibinfo {title}
  {{Q}uadrature formula for the sphere of $131$-th algebraic order of
  accuracy},}\ }\href@noop {} {\bibfield  {journal} {\bibinfo  {journal}
  {Doklady Mathematics}\ }\textbf {\bibinfo {volume} {59}},\ \bibinfo {pages}
  {477--481} (\bibinfo {year} {1999})}\BibitemShut {NoStop}%
\bibitem [{\citenamefont {Becke}(1988)}]{BeckeAD:1987JCP}%
  \BibitemOpen
  \bibfield  {author} {\bibinfo {author} {\bibfnamefont {A.~D.}\ \bibnamefont
  {Becke}},\ }\bibfield  {title} {\enquote {\bibinfo {title} {A multicenter
  numerical integration scheme for polyatomic molecules},}\ }\href@noop {}
  {\bibfield  {journal} {\bibinfo  {journal} {J. Chem. Phys.}\ }\textbf
  {\bibinfo {volume} {88}},\ \bibinfo {pages} {2547--2553} (\bibinfo {year}
  {1988})}\BibitemShut {NoStop}%
\bibitem [{\citenamefont {Groot-Berning}\ \emph {et~al.}(2014)\citenamefont
  {Groot-Berning}, \citenamefont {Raatz}, \citenamefont {Dobrinets},
  \citenamefont {Lesik}, \citenamefont {Spinicelli}, \citenamefont {Tallaire},
  \citenamefont {Achard}, \citenamefont {Jacques}, \citenamefont {Roch},
  \citenamefont {Zaitsev}, \citenamefont {Meijer},\ and\ \citenamefont
  {Pezzagna}}]{Groot2014}%
  \BibitemOpen
  \bibfield  {author} {\bibinfo {author} {\bibfnamefont {Karin}\ \bibnamefont
  {Groot-Berning}}, \bibinfo {author} {\bibfnamefont {Nicole}\ \bibnamefont
  {Raatz}}, \bibinfo {author} {\bibfnamefont {Inga}\ \bibnamefont {Dobrinets}},
  \bibinfo {author} {\bibfnamefont {Margarita}\ \bibnamefont {Lesik}}, \bibinfo
  {author} {\bibfnamefont {Piernicola}\ \bibnamefont {Spinicelli}}, \bibinfo
  {author} {\bibfnamefont {Alexandre}\ \bibnamefont {Tallaire}}, \bibinfo
  {author} {\bibfnamefont {Jocelyn}\ \bibnamefont {Achard}}, \bibinfo {author}
  {\bibfnamefont {Vincent}\ \bibnamefont {Jacques}}, \bibinfo {author}
  {\bibfnamefont {Jean-Fran{\c{c}}ois}\ \bibnamefont {Roch}}, \bibinfo {author}
  {\bibfnamefont {Alexander~M.}\ \bibnamefont {Zaitsev}}, \bibinfo {author}
  {\bibfnamefont {Jan}\ \bibnamefont {Meijer}}, \ and\ \bibinfo {author}
  {\bibfnamefont {S\'{e}bastien}\ \bibnamefont {Pezzagna}},\ }\bibfield
  {title} {\enquote {\bibinfo {title} {Passive charge state control of
  nitrogen-vacancy centres in diamond using phosphorous and boron doping},}\
  }\href@noop {} {\bibfield  {journal} {\bibinfo  {journal} {Phys. Status
  Solidi (a)}\ }\textbf {\bibinfo {volume} {211}},\ \bibinfo {pages}
  {2268--2273} (\bibinfo {year} {2014})}\BibitemShut {NoStop}%
\bibitem [{\citenamefont {Janssens}\ \emph {et~al.}(2011)\citenamefont
  {Janssens}, \citenamefont {Pobedinskas}, \citenamefont {Vacik}, \citenamefont
  {Petr{\'a}kov{\'a}}, \citenamefont {Ruttens}, \citenamefont {D'Haen},
  \citenamefont {Nesl{\'a}dek}, \citenamefont {Haenen},\ and\ \citenamefont
  {Wagner}}]{Janssens2011}%
  \BibitemOpen
  \bibfield  {author} {\bibinfo {author} {\bibfnamefont {SD}~\bibnamefont
  {Janssens}}, \bibinfo {author} {\bibfnamefont {Paulius}\ \bibnamefont
  {Pobedinskas}}, \bibinfo {author} {\bibfnamefont {J}~\bibnamefont {Vacik}},
  \bibinfo {author} {\bibfnamefont {V}~\bibnamefont {Petr{\'a}kov{\'a}}},
  \bibinfo {author} {\bibfnamefont {Bart}\ \bibnamefont {Ruttens}}, \bibinfo
  {author} {\bibfnamefont {Jan}\ \bibnamefont {D'Haen}}, \bibinfo {author}
  {\bibfnamefont {Milos}\ \bibnamefont {Nesl{\'a}dek}}, \bibinfo {author}
  {\bibfnamefont {Ken}\ \bibnamefont {Haenen}}, \ and\ \bibinfo {author}
  {\bibfnamefont {Patrick}\ \bibnamefont {Wagner}},\ }\bibfield  {title}
  {\enquote {\bibinfo {title} {Separation of intra-and intergranular
  magnetotransport properties in nanocrystalline diamond films on the metallic
  side of the metal--insulator transition},}\ }\href@noop {} {\bibfield
  {journal} {\bibinfo  {journal} {New J. Phys.}\ }\textbf {\bibinfo {volume}
  {13}},\ \bibinfo {pages} {083008} (\bibinfo {year} {2011})}\BibitemShut
  {NoStop}%
\bibitem [{\citenamefont {Drijkoningen}\ \emph {et~al.}(2016)\citenamefont
  {Drijkoningen}, \citenamefont {Janssens}, \citenamefont {Pobedinskas},
  \citenamefont {Koizumi}, \citenamefont {Van~Bael},\ and\ \citenamefont
  {Haenen}}]{Drijkoningen2016}%
  \BibitemOpen
  \bibfield  {author} {\bibinfo {author} {\bibfnamefont {Sien}\ \bibnamefont
  {Drijkoningen}}, \bibinfo {author} {\bibfnamefont {SD}~\bibnamefont
  {Janssens}}, \bibinfo {author} {\bibfnamefont {Paulius}\ \bibnamefont
  {Pobedinskas}}, \bibinfo {author} {\bibfnamefont {S}~\bibnamefont {Koizumi}},
  \bibinfo {author} {\bibfnamefont {MK}~\bibnamefont {Van~Bael}}, \ and\
  \bibinfo {author} {\bibfnamefont {Ken}\ \bibnamefont {Haenen}},\ }\bibfield
  {title} {\enquote {\bibinfo {title} {The pressure sensitivity of wrinkled
  b-doped nanocrystalline diamond membranes},}\ }\href@noop {} {\bibfield
  {journal} {\bibinfo  {journal} {Sci. Rep.}\ }\textbf {\bibinfo {volume}
  {6}},\ \bibinfo {pages} {35667} (\bibinfo {year} {2016})}\BibitemShut
  {NoStop}%
\bibitem [{\citenamefont {Gaillard}\ \emph {et~al.}(2013)\citenamefont
  {Gaillard}, \citenamefont {Adumeau}, \citenamefont {Canet}, \citenamefont
  {Gautier}, \citenamefont {Boyer}, \citenamefont {Beaudoin}, \citenamefont
  {Hesling}, \citenamefont {Morel},\ and\ \citenamefont
  {Mahiou}}]{Gaillard2013}%
  \BibitemOpen
  \bibfield  {author} {\bibinfo {author} {\bibfnamefont {Claire}\ \bibnamefont
  {Gaillard}}, \bibinfo {author} {\bibfnamefont {Pierre}\ \bibnamefont
  {Adumeau}}, \bibinfo {author} {\bibfnamefont {Jean-Louis}\ \bibnamefont
  {Canet}}, \bibinfo {author} {\bibfnamefont {Arnaud}\ \bibnamefont {Gautier}},
  \bibinfo {author} {\bibfnamefont {Damien}\ \bibnamefont {Boyer}}, \bibinfo
  {author} {\bibfnamefont {Claude}\ \bibnamefont {Beaudoin}}, \bibinfo {author}
  {\bibfnamefont {C{\'e}dric}\ \bibnamefont {Hesling}}, \bibinfo {author}
  {\bibfnamefont {Laurent}\ \bibnamefont {Morel}}, \ and\ \bibinfo {author}
  {\bibfnamefont {Rachid}\ \bibnamefont {Mahiou}},\ }\bibfield  {title}
  {\enquote {\bibinfo {title} {Monodisperse silica nanoparticles doped with
  dipicolinic acid-based luminescent lanthanide (iii) complexes for
  bio-labelling},}\ }\href@noop {} {\bibfield  {journal} {\bibinfo  {journal}
  {J. Mater. Chem. B}\ }\textbf {\bibinfo {volume} {1}},\ \bibinfo {pages}
  {4306--4312} (\bibinfo {year} {2013})}\BibitemShut {NoStop}%
\bibitem [{\citenamefont {Vanpoucke}\ \emph {et~al.}(2014)\citenamefont
  {Vanpoucke}, \citenamefont {Jaeken}, \citenamefont {Baerdemacker},
  \citenamefont {Lejaeghere},\ and\ \citenamefont
  {Speybroeck}}]{VanpouckeDannyEP:2014e_Beilstein}%
  \BibitemOpen
  \bibfield  {author} {\bibinfo {author} {\bibfnamefont {Danny E.~P.}\
  \bibnamefont {Vanpoucke}}, \bibinfo {author} {\bibfnamefont {Jan~W.}\
  \bibnamefont {Jaeken}}, \bibinfo {author} {\bibfnamefont {Stijn~De}\
  \bibnamefont {Baerdemacker}}, \bibinfo {author} {\bibfnamefont {Kurt}\
  \bibnamefont {Lejaeghere}}, \ and\ \bibinfo {author} {\bibfnamefont
  {Veronique~Van}\ \bibnamefont {Speybroeck}},\ }\bibfield  {title} {\enquote
  {\bibinfo {title} {Quasi-{1D} physics in metal-organic frameworks:
  {MIL-47(V)} from first principles},}\ }\href {\doibase 10.3762/bjnano.5.184}
  {\bibfield  {journal} {\bibinfo  {journal} {Beilstein J. Nanotechnol.}\
  }\textbf {\bibinfo {volume} {5}},\ \bibinfo {pages} {1738--1748} (\bibinfo
  {year} {2014})}\BibitemShut {NoStop}%
\bibitem [{fn:({\natexlab{b}})}]{fn:unphys}%
  \BibitemOpen
  \href@noop {} {} ({\natexlab{b}}),\ \bibinfo {note} {in these systems, the
  bottom of the conduction band of one spin component is located below the top
  of the valence band of the other spin component.}\BibitemShut {Stop}%
\bibitem [{fn:({\natexlab{c}})}]{fn:atomsize}%
  \BibitemOpen
  \href@noop {} {} ({\natexlab{c}}),\ \bibinfo {note} {the difference in atomic
  radius between C and Eu depends on the definition of atomic radius used: The
  empirical radius ($R_{C}= 70$ pm; $R_{Eu}=185$ pm; $\frac{R_{Eu}}{R_C} =
  2.64$), the covalent radius for single bonds ($R_{C}= 75$ pm; $R_{Eu}=168$
  pm; $\frac{R_{Eu}}{R_C} = 2.24$), the calculated radius ($R_{C}= 67$ pm;
  $R_{Eu}=231$ pm; $\frac{R_{Eu}}{R_C} = 3.45$), or the Shannon crystal radius
  ($R_{C}= 29$ pm; $R_{Eu}=108.7$ pm; $\frac{R_{Eu}}{R_C} =
  3.75$).}\BibitemShut {Stop}%
\bibitem [{\citenamefont {Vanpoucke}\ \emph {et~al.}(2012)\citenamefont
  {Vanpoucke}, \citenamefont {Cottenier}, \citenamefont {Van~Speybroeck},
  \citenamefont {Bultinck},\ and\ \citenamefont
  {Van~Driessche}}]{VanpouckeDannyEP:2012aApplSurfSci}%
  \BibitemOpen
  \bibfield  {author} {\bibinfo {author} {\bibfnamefont {Danny~E.~P.}\
  \bibnamefont {Vanpoucke}}, \bibinfo {author} {\bibfnamefont {Stefaan}\
  \bibnamefont {Cottenier}}, \bibinfo {author} {\bibfnamefont {Veronique}\
  \bibnamefont {Van~Speybroeck}}, \bibinfo {author} {\bibfnamefont {Patrick}\
  \bibnamefont {Bultinck}}, \ and\ \bibinfo {author} {\bibfnamefont {Isabel}\
  \bibnamefont {Van~Driessche}},\ }\bibfield  {title} {\enquote {\bibinfo
  {title} {Tuning of {CeO}$_2$ buffer layers for coated superconductors through
  doping},}\ }\href {\doibase 10.1016/j.apsusc.2012.01.032} {\bibfield
  {journal} {\bibinfo  {journal} {Appl. Surf. Sci.}\ }\textbf {\bibinfo
  {volume} {260}},\ \bibinfo {pages} {32--35} (\bibinfo {year}
  {2012})}\BibitemShut {NoStop}%
\bibitem [{\citenamefont {Subramanian}\ and\ \citenamefont
  {Han}(2013)}]{SubramanianH:JPhysCondMatter2013}%
  \BibitemOpen
  \bibfield  {author} {\bibinfo {author} {\bibfnamefont {Hemachander}\
  \bibnamefont {Subramanian}}\ and\ \bibinfo {author} {\bibfnamefont {J.~E.}\
  \bibnamefont {Han}},\ }\bibfield  {title} {\enquote {\bibinfo {title}
  {In-plane uniaxial magnetic anisotropy in {(Ga, Mn)As} due to local lattice
  distortions around {Mn} 2+ ions},}\ }\href
  {http://stacks.iop.org/0953-8984/25/i=20/a=206005} {\bibfield  {journal}
  {\bibinfo  {journal} {J. Phys.: Cond. Matter}\ }\textbf {\bibinfo {volume}
  {25}},\ \bibinfo {pages} {206005} (\bibinfo {year} {2013})}\BibitemShut
  {NoStop}%
\bibitem [{\citenamefont {Shannon}(1976)}]{Shannon:ACSA1976}%
  \BibitemOpen
  \bibfield  {author} {\bibinfo {author} {\bibfnamefont {R.~D.}\ \bibnamefont
  {Shannon}},\ }\bibfield  {title} {\enquote {\bibinfo {title} {{Revised
  effective ionic radii and systematic studies of interatomic distances in
  halides and chalcogenides}},}\ }\href {\doibase 10.1107/S0567739476001551}
  {\bibfield  {journal} {\bibinfo  {journal} {Acta Cryst.}\ }\textbf {\bibinfo
  {volume} {A32}},\ \bibinfo {pages} {751--767} (\bibinfo {year}
  {1976})}\BibitemShut {NoStop}%
\bibitem [{fn:({\natexlab{d}})}]{fn:visible}%
  \BibitemOpen
  \href@noop {} {} ({\natexlab{d}}),\ \bibinfo {note} {eu$^{2+}$ may also form
  luminescent centres, but their near-UV luminescence would have been outside
  of the detection window of our optical setup.}\BibitemShut {Stop}%
\bibitem [{fn:({\natexlab{e}})}]{fn:DOS}%
  \BibitemOpen
  \href@noop {} {} ({\natexlab{e}}),\ \bibinfo {note} {we did not include the
  DOS for the Eu$_{sub}^{7a}$ system as it was not the ground state
  configuration of the spinstate with 7 unpaired electrons. Similarly, we did
  not select the Eu$_{2V}^{7p}$ system, as it was not the groundstate of the
  hybrid functional calculations.}\BibitemShut {Stop}%
\end{thebibliography}%

\end{document}